\documentclass[preprint]{aastex}

\usepackage[numberedappendix]{emulateapj5}

\usepackage{amstext}
\usepackage{apjfonts}

\newcommand{\B}{{\mathbf{B}}}

\newcommand{\Ve}{{\mathbf{V}}}
\newcommand{\E}{{\mathbf{E}}}
\newcommand{\tni}{{\tau_{\text{ni}}}}
\newcommand{\tbr}{{\tau_{\text{br}}}}
\newcommand{\Bref}{{B_{\text{ref}}}}
\newcommand{\VAext}{{V_{\text{A,ext}}}}
\newcommand{\lfrac}[2]{{{#1}{/}{#2}}}

\newcommand{\second}{{\text{s}}}
\newcommand{\kelvin}{{\text{K}}}
\newcommand{\radian}{{\text{rad}}}
\newcommand{\gram}{{\text{g}}}
\newcommand{\gauss}{{\text{G}}}
\newcommand{\microgauss}{{\mu\text{G}}}
\newcommand{\cm}{{\text{cm}}}
\newcommand{\yr}{{\text{yr}}}
\newcommand{\km}{{\text{km}}}
\newcommand{\au}{{\text{AU}}}
\newcommand{\angstrom}{{\text{\AA}}}
\newcommand{\pc}{{\text{pc}}}

\newcommand{\kms}{{\km\,\second^{-1}}}
\newcommand{\blo}{{b_{z,\text{low}}}}
\newcommand{\bhi}{{b_{z,\text{high}}}}
\newcommand{\jplateau}{{j_{\text{pl}}}}
\newcommand{\mplateau}{{m_{\text{pl}}}}
\newcommand{\xplateau}{{x_{\text{pl}}}}
\newcommand{\QToomre}{{Q_{\text{Toomre}}}}
\newcommand{\range}{{\text{--}}}
\newcommand{\rin}{{r_{\text{in}}}}
\newcommand{\rout}{{r_{\text{out}}}}
\newcommand{\xmax}{{x_{\text{max}}}}
\newcommand{\mubar}{{{\bar{\mu}}}}
\newcommand{\etaWKL}{{\eta_{\text{WKL}}}}
\newcommand{\jconst}{{j_1}}

\def\emulapj@ver{11/12/01}

\begin{document}

\slugcomment{Submitted 2002 May 1; accepted 2002 August 5}

\journalinfo{The Astrophysical Journal, {\rm in press (Vol.\ 580, 2002 December 1)}}

\shorttitle{COLLAPSE OF ROTATING MAGNETIC CLOUD CORES}

\shortauthors{{KRASNOPOLSKY} AND {K\"ONIGL}}

\title{Self-similar Collapse of Rotating Magnetic Molecular Cloud Cores}

\author{Ruben Krasnopolsky and Arieh K\"onigl\altaffilmark{1}}
\affil{Department of Astronomy and Astrophysics, University of
Chicago, 5640 South Ellis Avenue, Chicago, IL 60637\\
\email{{\rm\{}ruben,arieh{\rm\}}@oddjob.uchicago.edu}}
\altaffiltext{1}{Also at the Enrico Fermi Institute, University
of Chicago.}

\begin{abstract}
We present self-similar solutions that describe the
gravitational collapse of rotating, isothermal, magnetic
molecular-cloud cores. These solutions make it
possible, for the first time, to study the formation of
rotationally supported protostellar disks of the type detected around
many young
stellar objects in the context of a realistic scenario of star formation
in magnetically supported, weakly ionized, molecular
cloud cores. This work focuses on the evolution
after a point mass first forms at the center and generalizes
previous results by Contopoulos, Ciolek, \& K\"onigl that did
not include rotation. Our semianalytic scheme incorporates
ambipolar diffusion and
magnetic braking and allows us to examine the
full range of expected behaviors and their dependence on the
physical parameters. We find that, for typical parameter values,
the inflow first passes through an ambipolar-diffusion shock (at
a radius $r_a$), where the magnetic flux decouples from the matter, and
subsequently through a centrifugal shock (at $r_c$), inward of
which a rotationally supported disk (of mass $M_d$) is
established. By the time ($\sim 10^5\, \yr$) that the central
mass $M_c$ grows to $\sim 1\, M_\odot$, $r_a \gtrsim 10^3\, \au$, $r_c
\gtrsim 10^2\, \au$, and $M_d/M_c \lesssim 0.1$. The derived
disk properties are consistent with
data on T Tauri systems, and our results imply that
protostellar disks may well be Keplerian also during
earlier phases of their evolution. We demonstrate that the disk
is likely to drive centrifugal outflows that transport angular
momentum and mass, and we show how the
radially self-similar wind solution of Blandford \& Payne
can be naturally incorporated into the disk model.
We further verify that gravitational torques and magnetorotational
instability-induced turbulence typically do not play an
important role in the angular momentum transport.
For completeness, we also present solutions for the
limiting cases of fast rotation (where the collapse results in a massive
disk with such a large outer radius that it traps
the ambipolar-diffusion front) and strong braking (where
no disk is formed and the collapse resembles that of a nonrotating core at
small radii), as well as
solutions describing the rotational collapse of ideal-MHD and of
nonmagnetic model cores.
\end{abstract}

\section{Introduction}
\label{introduction}

Low-mass star formation is believed to occur predominantly in
molecular cloud cores. The commonly accepted scenario is that
dense molecular clouds are supported against self-gravity by
magnetic and thermal stresses. Since the gas is only weakly
ionized, the clouds are not in a strict steady state: the prevalent
neutral molecules, pulled in by the force of gravity, gradually
drift inward through the magnetic field (attached to the ionized
component and anchored in the cloud envelope) in a process known
as ambipolar diffusion. When the mass-to-flux ratio in the cloud core exceeds a
critical value, the core undergoes gravitational collapse to
form a central protostar. The core evolution after the collapse
is initiated can in
general be divided into two phases: a rapid initial dynamical
contraction, which proceeds nonhomologously and eventually
results in the formation of a central point mass, followed by an
accretion phase, during which the central mass gradually
increases and its gravitational field dominates a progressively
larger region around the origin. The evolution of star-forming molecular
clouds has been studied mostly through numerical simulations
(see, e.g., \citealt{MC1999} for a review). Until recently, these
calculations were terminated before the end of the
dynamical-collapse phase, basically at the
point where radiative trapping near the center started to invalidate
the isothermality assumption adopted in these
computations. However, \citeauthor{CK98} (\citeyear{CK98}, hereafter CK)
circumvented this problem by treating the comparatively small
region of radiative trapping as a central sink cell; this
enabled them to model the collapse of nonrotating magnetic cloud cores
through point-mass formation (PMF) and into the protostellar accretion phase.
Concurrently, \citeauthor{CCK98} (\citeyear{CCK98}, hereafter CCK)
obtained semianalytic self-similar solutions that explicitly
accounted for the effects of ambipolar diffusion and
successfully reproduced the main qualitative features of the
numerical simulations of CK.

Our goal in this paper is to extend the previous work on the
post-PMF evolution of collapsing magnetic cloud cores by
including the effects of rotation. Molecular line observations
(e.g., \citealt{Goodman1993}; \citealt{Kane1997})
have established
that a majority of dense ($\gtrsim 10^4\, \cm^{-3}$)
cloud cores show evidence of rotation, with angular velocities\linebreak
$\sim 3{\times}10^{-15}{\range}10^{-13}\, \second^{-1}$ that tend
to be uniform on scales of $\sim 0.1\, \pc$, and
with specific angular momenta in the range\linebreak
$\sim 4{\times}10^{20}{\range}3{\times}10^{22}\, \cm^2\,\second^{-1}$.
Although present, the rotation
contributes only a small fraction (typically no more than a few
percent) to the dynamical support of these cores, and self-gravity
is mostly balanced by magnetic and thermal stresses. During
their quasistatic contraction phase, the cores evidently lose angular
momentum by means of magnetic braking --- the magnetic transfer of
angular momentum to the ambient gas through torsional Alfv\'en
waves --- and this process also tends to align their angular
velocity and large-scale magnetic field vectors \citep[e.g.,][]{MC1999}.
Once dynamical collapse is initiated and a core goes into a
near--free-fall state, the
specific angular momentum is expected to be approximately
conserved, resulting in a progressive increase in the
centrifugal force that eventually halts the collapse and gives
rise to a rotationally supported disk. These expectations are
consistent with the results of molecular-line interferometric
observations, which have probed contracting cloud cores on scales
$\sim 10^2{\range}10^3\, \au$ \citep*[e.g.,][]{Myers2000,Mundy2000,Wilner2000}.
These observations have revealed that,
on scales $\lesssim 10^3\, \au$, the cores have a flattened,
thick-disk morphology and a velocity field that is dominated by
infall motions (with only a modest rotational component). This
morphology is consistent with numerical simulations of
magnetically supported clouds, which have demonstrated that the
gas rapidly contracts along the field lines and maintains force
equilibrium along the field even during the collapse phase (e.g.,
\citealt{FM1993}; \citealt{GalliShu1993}; CK),
including in cases where the clouds
are initially elongated in the field direction
\citep[e.g.,][]{Nakamura1995,Tomisaka1996}. 
The observations have also established
that angular momentum is by and large conserved in the infalling
gas \citep[e.g.,][]{Ohashi1997} and leads to the formation of
rotationally supported disks on scales $\lesssim 10^2\, \au$.

The pre-PMF collapse of rotating, magnetized, weakly ionized cloud cores was
previously studied by \citet{BM1994,BM1995a,BM1995b}. They
found that little angular momentum is lost during the
dynamical-collapse phase but that the centrifugal force nevertheless does
not become important prior to PMF. In the current work we focus on the post-PMF
phase of the collapse, which enables us to model the formation
of circumstellar disks in the context of magnetic cloud cores
and to study their properties and their role in the buildup of the
central protostar.
Circumstellar disks have been detected in
$\sim 25{\range}50\%$ of pre--main-sequence stars in nearby dark clouds
\citep[e.g.,][]{Beckwith1993}, and it is likely that most of
the mass assembled in a typical low-mass young stellar object
(YSO) is accreted through a disk \citep*[e.g.,][]{Calvet2000}.
The study of protostellar disks is important also
because they are the incubators of planetary systems, so their
properties are directly relevant to the process of planet formation
\citep[e.g.,][]{Wuchterl2000}. The rotational properties of the
collapsing core are evidently also of particular relevance to
the question of multiple-star formation
\citep[e.g.,][]{Bodenheimer2000}. In the ensuing
discussion we assume that the collapsing core does not give rise to more
than one stellar object; we nevertheless briefly address the
issue of core fragmentation in \S~\ref{discussion}.

The studies of CK and CCK demonstrated that ambipolar
diffusion, which is unimportant during the dynamical pre-PMF
collapse phase, is ``revitalized'' after a point mass starts to
grow in the center and leads to a decoupling of the magnetic
flux from the inflowing gas. It was found that the ``decoupling
front'' propagates outward in the form of a hydromagnetic shock
(as originally suggested by \citealt{LiMcKee1996}) and that this
process can go a long way toward resolving the magnetic flux
problem in star formation (the several-orders-of-magnitude
discrepancy between the empirical upper limit on the magnetic
flux of a protostar and the flux associated with the
corresponding mass in the pre-collapse core). By incorporating
rotation into the picture, we can study how this issue is
affected by the formation of a disk. Furthermore, as our model
includes angular momentum transport (by the large-scale magnetic field),
which is crucial for enabling mass to accumulate in the
center and form a protostar (see \S~\ref{formulation}), we are
also in a position to address the angular momentum problem in star
formation (analogous to the magnetic flux problem, except that
the quantity under consideration is the specific angular momentum).

The emergence of a quasi-stationary, rotationally supported disk from
a nearly freely falling, collapsing core inevitably involves a strong
deceleration of the inflowing matter, which almost invariably
takes place in a shock. This ``centrifugal'' shock is distinct from
the ambipolar-diffusion
shock mentioned above: it is typically hydrodynamic,
rather than hydromagnetic, in nature, and it is generally
located at a different distance from the center. The appearance
of a second shock increases the complexity of the problem, and
it might seem that it could only be tackled through a
multi-scale numerical simulation. However, we have found that it
is possible to treat this problem semianalytically by
generalizing the self-similar solutions presented in CCK to
include the effects of rotation and magnetic braking. Although
this treatment is somewhat less detailed than
a numerical simulation, the semianalytic approach allows us to
more readily explore the fairly extensive parameter
space. This approach is supported by the strong evidence from numerical
simulations that a multiscale core collapse naturally leads to a
self-similar evolution. In particular, this behavior has been
found to describe the effectively magnetic flux- and angular
momentum-conserving {\em pre}-PMF collapse phase of a rotating magnetic core
seen in the aforementioned numerical simulations of
\citeauthor{BM1994} (\citealt{Basu1997}; \citealt{Nakamura1999}; see also
\citealt*{Narita1984} and \citealt{LiShu1997}). As regards the
post-PMF evolution, \citet{SH} obtained
similarity solutions for flattened cloud cores that approximate
previous numerical simulations of the collapse of rotating clouds
through the moment when a central point mass could be expected
to form. The good
correspondence between the CCK solutions and the nonrotating magnetic collapse
simulations of CK further strengthens our confidence in the
viability of this approach.

In relating this work to previously published results, we note
that neither the \citet{SH} model nor
the numerical simulations with which they compare their
solutions incorporate an angular momentum transport mechanism,
and therefore they do not in practice give rise to a central
point mass. (Nevertheless, as we discuss in \S~\ref{nonmagnetic}, their
post-PMF solutions correspond to a limiting case of the more general
model considered in this paper.) It is also worth mentioning in
this connection that self-similar models of the gravitational
collapse of rotating, disk-like mass distributions that
incorporate {\em viscous} angular momentum transport have been
constructed by \citet{Mineshige} and \citet{Tsuribe}.
However, the former model arbitrarily imposes an inner
boundary condition that does not allow mass to accumulate at the
center, whereas the latter model is restricted to the case in which
the inflow speed is everywhere so low that no centrifugal shock forms.
Neither of these models therefore provides a realistic framework
for studying star formation in rotating molecular cloud cores.

The plan of this paper is as follows. In \S~\ref{model} we
formulate the problem, discuss the adopted approximations, and
describe the self-similar model. In \S~\ref{results} we present
illustrative solutions, ordered by increasing level of complexity, which
isolate different aspects of the full problem. We use the
flexibility afforded by our semianalytic approach to examine the
dependence of the results on the efficiency of ambipolar
diffusion and magnetic braking and to elucidate the interplay
between these two processes. We give a general discussion of
these solutions and of their astrophysical implications in
\S~\ref{discussion}, where we also consider the production of
centrifugally driven disk outflows.
Our conclusions are summarized in \S~\ref{summary}.

\section{Self-Similar Model}
\label{model}

\subsection{Formulation of the Problem}
\label{formulation}

Our basic model setup is the same as in CCK. We consider
an isothermal, disklike core that contracts in the radial
direction while maintaining a vertical hydrostatic equilibrium at all times.
As noted in \S~\ref{introduction}, this behavior was found in
numerical simulations of isothermal cores that are initially
supported against gravitational collapse by a large-scale,
ordered magnetic field. Nonmagnetic simulations
\citep*[e.g.,][]{Norman1980,Narita1984,Matsumoto1997}
have established that
rotation could further contribute to the tendency of a core to
flatten as it contracts so long as the specific angular momentum
is approximately conserved. In restricting gas motions in our
model to the plane of the flattened core, we preclude the
possibility that matter can reach the center from above (or
below) the disk, and this, in turn, implies that a point mass
can form in the center only if there exists a mechanism for
removing angular momentum from the inflowing gas. There have
also been models in which the complementary view has been
adopted, viz., new mass is added to the protostar and its
circumstellar disk only from above and below the disk plane
\citep[e.g.,][]{Terebey1984,Yorke1993}.
These models tend to produce rotationally
supported disks whose masses are comparable to that of the
central star, even if angular momentum transport within the disk
is taken into account \citep[e.g.,][]{Cassen1983,LinPringle1990}.
However, most observed protostellar disks in
nearby dark clouds have masses (estimated from dust
emission or, when possible, from a direct measurement of the
disk velocity field) that are significantly smaller ($\lesssim 10\%$)
than the protostellar mass \citep[e.g.,][]{Beckwith1993}. As
we demonstrate in \S~\ref{results}, such low-mass systems can form naturally
when mass flows into the central region primarily in the disk
plane. Given that this is the expected situation for magnetically
regulated core contraction and that the disklike configurations
predicted in this case are consistent with observations of
collapsing cloud cores, we consider this approximation to be
adequate for capturing the essence of protostellar-disk and star formation.

The isothermality assumption was justified in the work of CK
and CCK on the grounds that it generally applies so long as the
density remains below $\sim 10^{10}\, \cm^{-3}$, which for a
typical nonrotating collapse is the case on scales $\gtrsim 5\, \au$.
This assumption is less tenable in the present work because,
when a quasi-stationary, centrifugally supported disk forms, it
has a significantly higher column density at a given radius
than a nearly freely falling configuration, with the result that
radiative trapping already occurs on scales $\lesssim 10^2\, \au$
\citep[e.g.,][]{WK}. Irradiation by the central
protostar tends to mitigate this effect and establish vertical
isothermality in the outer regions of the disk, but, in turn, it
leads to a moderately strong ($T \propto r^{-1/2}$) variation of
the disk temperature with radius \citep[e.g.,][]{DAlessio1998}.
However, since thermal stresses turn out not to
play a major role in the core-collapse dynamics, the assumption
of isothermality (which is required for self-similarity) is not
likely to introduce significant inaccuracies into the results.

Another requirement for self-similarity is that the ion density
$\rho_i$ be proportional to the square root of the neutral mass
density (which is essentially equal to the total gas density
$\rho$). In this case, the
ratio $\eta$ of the neutral--ion momentum-exchange time scale
$\tni$ to the nominal self-gravity free-fall time
$(4\pi G \rho)^{-1/2}$ is independent of density and can be treated as a
constant.\footnote{Our dimensionless ambipolar diffusion parameter
$\eta\equiv\tni\left(4\pi G\rho\right)^{1/2}$ is smaller than
the one used in CCK by a factor $\sqrt{2}$.}
This behavior is found at comparatively low densities,
when molecular ionization by cosmic rays (at a rate $\xi =
10^{-17}\xi_{-17}\, \second^{-1}$) is balanced by rapid dissociative
recombination of the molecular ions and the dominant positive charges
are metal ions (formed by a charge-exchange reaction
between molecular ions and neutral metal atoms and destroyed by
recombination on grain surfaces). For grains of radius
$a=0.1\, \micron$, a temperature $T=10\, \kelvin$, and $\xi_{-17}
\approx 1$, this scaling applies roughly at neutral particle densities between
$\sim 10^4\, \cm^{-3}$ and $\lesssim 10^7\, \cm^{-3}$
\citep[e.g.,][]{CM1998,KN2000}.
These values of $T$ and
$n$ characterize the outer regions of collapsing cores,
typically on scales $\gtrsim 10^3\, \au$ (e.g., CK). In this
regime, $\eta \approx 0.2\, \xi_{-17}^{-1/2}$.

The numerical models of CK extended to scales $\lesssim 10\, \au$,
where densities $\gtrsim 10^{10}\, \cm^{-3}$ were
attained. In the case of a rotationally supported disk, even
higher densities are expected on these scales. At such high
densities, grains are typically the dominant carriers of both
positive and negative charges, and their densities also
scale as $n^{1/2}$ \citep*[e.g.,][]{Nishi1991}.
Assuming, for the sake of illustration, that
the charged grain distribution is dominated by a population of small
(PAH-like) singly charged particles of radius $a$
\citep[see][]{Neufeld1994}, we can deduce the charged-grain
number density $n_\pm$ from the ionization balance relation
\begin{equation}\label{charge}
\xi n = \pi (2a)^2 n_\pm^2
\left ( \frac{8 k T}{\pi m_{\pm r}}\right )^{1/2}
\left ( 1 + \frac{e^2}{2 a k T} \right )\, ,
\end{equation}
where $m_{\pm r} = 2\pi a^3\rho_s/3$
(with $\rho_s \approx 2.3\, \gram\, \cm^{-3}$)
is the reduced grain mass. Taking as
fiducial values $n = 10^{12}n_{12}\, \cm^{-3}$ and $T =
10^2T_2\, \kelvin$ at a distance $r \approx 10\, \au$ from the
center of a rotationally supported protostellar disk, we can
neglect the factor 1 in comparison with the second term in the
last parentheses on the right-hand side of equation (\ref{charge})
(which accounts for the electrostatic attraction
between the recombining grains). We then get $n_\pm
\approx 15.8 (a/5\, \angstrom)^{1/4}T_2^{1/4}\xi_{-17}^{1/2}
n_{12}^{1/2}\, \cm^{-3}$. Assuming that the small grains are characterized by
the same rate coefficient for momentum transfer through elastic
scattering with neutrals as metal ions ($\left<\sigma v\right>_{\pm n}
\approx 1.7{\times}10^{-9}\, \cm^3\,\second^{-1}$; see
\citealt{WardleNg1999}), we
obtain $\eta \approx 0.07 \xi_{-17}^{-1/2}T_2^{-1/4}(a/5\, \angstrom)^{-1/4}$
in this density regime. Ambipolar diffusion will dominate Ohmic diffusivity
so long as the grain Hall parameter $\beta_\pm$ (the ratio of
the cyclotron frequency to the collision frequency with
neutrals) exceeds 1. Adopting a fiducial
magnetic field strength at $r = 10\, \au$ of $B = 0.3\, \gauss$,
we find $\beta_\pm \approx 0.7(B/0.3\, \gauss)/n_{12}$, which
we expect to climb securely above 1 as the density continues to decrease with
increasing $r$. We thus infer that both the low- and the
high-density regimes of our model collapsing cores correspond to
ambipolar diffusion that is amenable to self-similar
scaling. The intermediate density regime will depart from this
scaling \citep[e.g.,][]{Nishi1991}, and it is also likely to
involve significant contributions from the Hall current term in
Ohm's law \citep[e.g.,][]{WK,WardleNg1999}.
Nevertheless, given that our bracketing values of the
parameter $\eta$ vary by only one order of magnitude (assuming that
$\xi_{17}$ lies in the range $\sim 1{\range}10$) across $\sim 8$ orders
of magnitude in density, we may expect our diffusivity
parameterization to yield qualitatively meaningful results on all scales
between $\sim 10\, \au$ and $\sim 10^4\, \au$.

Just as in CCK, we take the initial ($t=0$) state of our model
core to coincide with the endpoint of the pre-PMF collapse phase,
which is characterized by a radial scaling of the surface
density $\Sigma$ and the vertical magnetic field component $B_z$
$\propto r^{-1}$ and by spatially constant radial
inflow velocity $V_r(t=0) \equiv u_0 C$ (where $C$ is the isothermal
speed of sound) and mass accretion rate $\dot
M_0 \equiv - A (u_0 C^3/G)$. In the presence of rotation, this state
also exhibits a uniform azimuthal speed. Since the latter is typically much
smaller than the speed of sound
$C \approx 0.19(T/10\, \kelvin)^{1/2}\, \kms$, we
normalize our initial state as in CCK based on the nonrotational collapse
simulations of CK ($u_0 \approx -1$ and $A \approx 3$). We
determine the initial value of $V_{\phi}/C$ (which we denote by
$v_0$) from the expression
\begin{eqnarray}
v_0 \approx
\frac{A \Omega_b C}{\sqrt{G} \Bref}\nonumber
=& & 0.15
\left ( \frac{A}{3} \right )
\left ( \frac{\Omega_b}{2{\times}10^{-14}\, \radian\,\second^{-1}} \right )
\times\\
&\times& \left ( \frac{C}{0.19\, \kms } \right )
\left ( \frac{\Bref}{30\, \microgauss} \right )^{-1}
\label{v0}
\end{eqnarray}
\citep[see][]{Basu1997}, where $\Omega_b$ and $\Bref$ are,
respectively, the uniform background angular velocity and
magnetic field strength. Based on the range of
measured core angular velocities (see \S~\ref{introduction}),
$v_0$ could be a factor $\sim 5$ larger or smaller than the
fiducial value adopted in equation (\ref{v0}).

We assume that magnetic braking, which regulates the core
rotation prior to its collapse, remains the dominant angular
momentum transport mechanism also during the subsequent
evolution.
[A centrifugally driven disk wind may, however,
dominate the angular momentum transport in the rotationally supported
disk. In \S~\ref{results} we argue that this is, indeed, a likely
possibility, and in Appendix~\ref{wind} we show that this
mechanism can be incorporated into
our model without significantly modifying the basic formulation.]
We adopt the approach articulated by \citet{BM1994}
for the pre-PMF collapse phase. The torque per
unit mass on the slab-like core (of surface mass density
$\Sigma$) can be approximated by $rB_zB_{\phi,s}/2\pi\Sigma$,
where $B_z$ is the vertical field component at the midplane
and $B_{\phi,s}$ is the azimuthal field component at the surface
$z=H$ (assuming $H\ll r$). The latter can be estimated from the
relation
\begin{equation}\label{BM}
B_{\phi,s}(r) =
-\frac{\Psi(r)}{\pi r^2}
\left(\frac{V_{i,\phi}(r) - r\Omega_b}{\VAext}\right)\ ,
\end{equation}
where $\Psi(r)$ is the magnetic flux threading the core within a
radius $r$, $V_{i,\phi}=r\Omega_i$ is the azimuthal speed of the core
ions, and $\VAext$ is the speed with which
the torsional Alfv\'en waves that effect the magnetic braking propagate
in the external medium. Equation (\ref{BM}) is the same as
equation (26) in \citet{BM1994}, where $\VAext$
is also treated as a constant, except that we
substituted the ion speed for the bulk (neutral) speed to take
account of the possibility that the ions (into which the magnetic
field is frozen) are not well coupled to the neutral core component
(see \citealt{MP1986} and \citealt{Konigl1987}).

The assumption of a constant $\VAext$ is consistent with
theoretical models of magnetically supported clouds as well as
with empirical data, which indicate that $\VAext\approx 1\, \kms$ in
molecular clouds over at least 4 orders of magnitude in density
($\sim 10^3{\range}10^7\, \cm^{-3}$; e.g.,
\citealt{Crutcher1999}). This assumption is required
for the construction of a self-similar model. In order for the
steady-state expression
(\ref{BM}) to be applicable, the Alfv\'en travel time in the
external medium over the initial radius of the cloud should be
less than the evolutionary time scale $\sim r/|V_r|$ (which, for
a rotationally supported disk, is essentially the local magnetic
braking time). Our representative rotationally supported disk
models typically have $|V_r| \lesssim 0.1 C$ (with $|V_r|
\rightarrow 0$ as $r\rightarrow 0$), whereas, for a nominal
temperature of $10\, \kelvin$, $\VAext/C \approx 5$. These
estimates indicate that the assumption of rapid braking should
not lead to gross errors even if it is not everywhere strictly
correct. In our models we use the parameterization $\VAext=
C/\alpha$, with $\alpha = 0.1$ adopted as a typical value.

We will be interested in evaluating the effects of magnetic
braking in regions where rotation already plays a dynamically
significant role, so that $\Omega_i \gg \Omega_b$. We can therefore
drop $\Omega_b$ in equation (\ref{BM}), which will make it
possible to incorporate this equation into a self-similar
formulation. The anticipated increase in $\Omega_i$ with
decreasing $r$ ($\propto r^{-3/2}$ when a Keplerian disk is
formed) will in general make $B_{\phi,s}$ the dominant
field component at the surface as the central point mass is
approached. (This situation is not encountered in the
pre-PMF phase; see \citealt{BM1994}.) We expect,
however, that various magnetohydrodynamic instabilities (in
particular, internal kinks) will intervene to prevent the azimuthal
field component from greatly exceeding the poloidal
components.\footnote{As we discuss in \S~\ref{MHDstrong}, we
also do not expect the radial surface field component to greatly
exceed the vertical field $B_z$ under realistic circumstances.}
We therefore impose a cap on the azimuthal field in the form
$\left|B_{\phi,s}\right| \leqslant \delta B_z$.
In our models we usually set $\delta = 1$ ---
this choice, in fact, also corresponds to the typical value of
$|B_{\phi,s}|/B_z$ obtained in our model for a rotationally
supported, diffusive disk when (as expected) the vertical angular momentum
transport is dominated by a centrifugally driven wind (see
\S~\ref{discussion} and Appendix~\ref{wind}).

\subsection{Basic Equations}
\label{basiceqns}
We represent the collapsing core as a thin disk surrounding a
point mass $M_c$. We use cylindrical coordinates $(r,\phi,z)$
centered on the point mass, with the disk midplane given by
$z=0$. The disk has a column density $\Sigma$, mass density $\rho$,
and half-thickness $H$ defined by $\Sigma=2H\rho$. The
total mass $M(r)$ enclosed within the radius $r$ is then given by
$M(r)=M_c+2\pi\int_0^r\Sigma(r') r'dr'$, and we denote its time
derivative by $\dot{M}$. The disk velocity field $\mathbf{V}(r)$
has radial and azimuthal components, with the latter giving rise
to a specific angular momentum $J=rV_\phi$. We assume that the
disk is threaded by an open magnetic field
that is symmetric with respect to the midplane [i.e.,
$B_r=0$ at $z=0$ and $B_{r,s}(H) = -B_{r,s}(-H)$ at the
disk surfaces ($z=\pm H$), and similarly for $B_{\phi}$].
The magnetic flux enclosed within the radius $r$ is given by
$\Psi(r)=\Psi_c+2\pi\int_0^rB_z(r')r'dr'$, where $\Psi_c$
denotes the flux trapped inside the central point mass.

In Appendix~\ref{diskequations} we write down the basic disk equations,
integrate them over $z$ while retaining all terms of order
$H/r$, and then go on to justify the omission of some of the
$\mathcal{O}(H/r)$ terms in the interest of greater
simplification. The result of these manipulations is the
following set of equations,
\begin{equation}
\label{massdimensional}
\frac{\partial\Sigma}{\partial t}
+
\frac{1}{r}\frac{\partial}{\partial r}\left(rV_r\Sigma\right)=0\ ,
\end{equation}
\begin{equation}
\label{forcedimensional}
\frac{\partial V_r}{\partial t}
+
V_r\frac{\partial V_r}{\partial r}=
g_r
-\frac{C^2}{\Sigma}\frac{\partial\Sigma}{\partial r}
+\frac{B_z}{2\pi\Sigma}\left(B_{r,s}-H\frac{\partial B_z}{\partial r}\right)
+\frac{J^2}{r^3}\ ,
\end{equation}
\begin{equation}
\label{angulardimensional}
\frac{\partial J}{\partial t}
+
V_r\frac{\partial J}{\partial r}=
\frac{rB_zB_{\phi,s}}{2\pi\Sigma}\ ,
\end{equation}
and
\begin{eqnarray}
\frac{\Sigma C^2}{2 H} &=& \frac{\pi}{2}G\Sigma^2
+ \frac{GM_c\rho H^2}{2r^3}\nonumber\\
& & \mbox{}+ \frac{1}{8\pi}\left(
B_{\phi,s}^2
+B_{r,s}^2-B_{r,s}\, H\frac{\partial B_z}{\partial r}
\right)\ ,
\end{eqnarray}
which express mass, radial momentum, and
angular momentum conservation, and vertical hydrostatic equilibrium,
respectively.

Under the assumption that the magnetic field evolution is governed
by ambipolar diffusion, one can regard the magnetic field lines
as being frozen into the ions. The latter move with velocity
$\mathbf{V}_i$ and drift with respect to the bulk (neutral) gas
component at the drift velocity $\mathbf{V}_D\equiv\Ve_i-\Ve$. By
balancing the Lorentz force on the ions with the ion--neutral
collisional drag force, one can solve for the components of $\mathbf{V}_D$:
\begin{eqnarray}
V_{D,r}&=&\frac{\tni B_z}{2\pi\Sigma}\left(
B_{r,s}-H\frac{\partial B_z}{\partial r}\right)\ ,\\
V_{D,\phi}&=&\frac{\tni B_zB_{\phi,s}}{2\pi\Sigma}\ .
\label{phidrift}
\end{eqnarray}
We then have
\begin{equation}
\frac{\partial\Psi}{\partial t}=
-2\pi r V_{i,r}B_z
=-2\pi r\left(V_r+V_{D,r}\right)B_z\ .
\end{equation}

As in CCK, we adopt the monopole approximations for $g_r$ and $B_{r,s}$,
\begin{eqnarray}
g_r&=&-\frac{GM(r,t)}{r^2}\ ,\\
B_{r,s}&=&\frac{\Psi(r,t)}{2\pi r^2}\ ,
\end{eqnarray}
which considerably simplify the calculations and are not
expected to introduce any significant errors
(see also \citealt{LiShu1997}, \citealt{SH}, CK, and \citealt{Tsuribe}).

For the azimuthal magnetic field component, we use equation (\ref{BM})
in the limit $V_{i,\phi}/r \gg \Omega_b$ together with
equation (\ref{phidrift}), and impose the bounding condition on $|B_{\phi,s}|$
discussed at the end of \S~\ref{formulation}. We then obtain
\begin{equation}
\label{bphis}
B_{\phi,s}=-\min\left[
\frac{\Psi}{\pi r^2}\!
\frac{V_\phi}{\VAext}
\left(
1+
\frac{\Psi B_z \tni}{2\pi^2r^2\Sigma\VAext}
\right)^{-1}\!;
\delta B_z\right]\,.
\end{equation}

\subsection{Self-Similar Equations in Nondimensional Form}
\label{selfsimilareqns}
We introduce a similarity variable $x$ and a set of
nondimensional flow quantities that depend only on $x$:
\begin{equation}
x=r/Ct\ ,
\end{equation}
\begin{equation}
H(r,t)=Ct\,h(x)
\, ,\ \ \Sigma(r,t)=(C/2\pi Gt)\,\sigma(x)\ ,
\end{equation}
\begin{equation}
V_r(r,t)=C\,u(x)
\, ,\ \ V_\phi(r,t)=C\,v(x)\ ,
\end{equation}
\begin{equation}
g_r(r,t)=(C/t)\,g(x)
\, ,\ \ J(r,t)=C^2t\,j(x)\ ,
\end{equation}
\begin{equation}
M(r,t)=(C^3t/G)\,m(x)\, ,
\ \ \dot{M}(r,t)=(C^3/G)\,\dot{m}(x)\ ,
\end{equation}
\begin{equation}
\B(r,t)=(C/G^{1/2}t)\,{\mathbf{b}}(x)\, ,
\Psi(r,t)=(2\pi C^3t/G^{1/2})\psi(x)\ .
\end{equation}
These definitions extend the set used in CCK to the rotational
case (which involves the dimensionless azimuthal speed $v$ and
specific angular momentum $j$) --- note, however, that the variables
$h$, $u$, $\sigma$, and $\mu\equiv\sigma/b_z$
in the present paper were denoted by
$\hat h$, $v$, $a$, and $\lambda$, respectively, in CCK.
To help relate our results to real protostellar systems, we
observe that, for our fiducial value of $C$ ($=0.19\, \kms$), $x=1$
corresponds to a distance $r\approx 6\times 10^{15}\, \cm$
($=400\, \au$) when $t=10^4\, \yr$
(the characteristic age of a Class 0 YSO) and to a distance
$r\approx 6\times 10^{16}\, \cm$ ($=4000\, \au$) when $t=10^5\, \yr$ (the
characteristic age of a Class I YSO).

CCK demonstrated that ambipolar diffusion can be incorporated into a
self-similar model when $\rho_i \propto \rho^{1/2}$. When
rotation is present, we find that magnetic braking can similarly
be included if $\VAext = {\text{const}}$. The nondimensional
model parameters that control the strength of these two effects
are $\eta$ and $\alpha$, respectively (see
\S~\ref{formulation}).
[A self-similar formulation is also possible if
angular momentum transport is due to an ``$\alpha$ viscosity,'' $\nu =
\alpha_{\text{SS}} C H$ \citep{SS1973}, when $C$ and the
parameter $\alpha_{\text{SS}}$ ($\lesssim 1$) are constant. This
was noted by \citet{Tsuribe} in the special case where $H$ has the
value appropriate to a Keplerian disk, and by \citet{Mineshige}
in the case when $H$ has the value appropriate to
a self-gravitating disk (although the latter authors required
$\alpha_{\text{SS}}H/r$,
rather than $\alpha_{\text{SS}}$, to be a constant).]
With the above expressions for the similarity variables, and defining also
$w\equiv x-u$ for convenience, the structure equations can be
rewritten in the following nondimensional, self-similar form:
\begin{equation}
\label{uequation}
\frac{du}{dx}=
w\left(\frac{1}{\sigma}\frac{d\sigma}{dx}+\frac{1}{x}\right)\ ,
\end{equation}
\begin{equation}
\label{forceequation}
\left(1-w^2\right)
\frac{1}{\sigma}\frac{d\sigma}{dx}=
g+\frac{b_z}{\sigma}
\left(b_{r,s}-h\frac{db_z}{dx}\right)
+\frac{j^2}{x^3}
+\frac{w^2}{x}\ ,
\end{equation}
\begin{equation}
\label{psiequation}
\psi-
xwb_z+\eta xb_z^2h^{1/2}\sigma^{-3/2}
\left(b_{r,s}-h\frac{db_z}{dx}\right)=0\ ,
\end{equation}
\begin{equation}\label{angmomequation}
\frac{dj}{dx}=w^{-1}\left(j-xb_zb_{\phi,s}/\sigma\right)\ ,
\end{equation}
\begin{equation}
\label{bphiequation}
b_{\phi,s}=-\min\left\{
\frac{2\alpha\psi j}{x^3}
\left(1+\frac{2\alpha\eta h^{1/2}\psi
b_z}{\sigma^{3/2}x^2}\right)^{-1}
\ ;\ \delta b_z\right\}\ ,
\end{equation}
\begin{equation}
\label{mw}
m=xw\sigma\, ,
\end{equation}
\begin{equation}
\label{mdot}
\dot{m}=-xu\sigma\, ,
\end{equation}
\begin{equation}
\label{msigma}
\frac{dm}{dx}=x\sigma\, ,
\end{equation}
\begin{equation}
\frac{d\psi}{dx}=xb_z\, ,
\end{equation}
\begin{equation}
g=-m/x^2\, ,
\end{equation}
and
\begin{equation}
b_{r,s}=\psi/x^2\, .
\end{equation}

The disk half-thickness $h$ can be found from the vertical
hydrostatic equilibrium condition, which yields
\begin{equation}
\label{thicknessequation}
\left(\frac{\sigma m_c}{x^3} - b_{r,s}\frac{d b_z}{d
x}\right)h^2+\left(b_{r,s}^2+b_{\phi,s}^2+\sigma^2\right)h -
2\sigma= 0\, .
\end{equation}
The solution of this quadratic equation is
\begin{equation}
\label{thicknesssolution}
h=\frac{\hat \sigma x^3}{2\hat m_c}
\left[-1 + \left(1+\frac{8\hat m_c}{x^3\hat \sigma^2}\right)^{1/2}\right]\ ,
\end{equation}
where $\hat m_c\equiv m_c-x^3b_{r,s}(db_z/dx)/\sigma$ and $\hat
\sigma \equiv \sigma +
(b_{r,s}^2+b_{\phi,s}^2)/\sigma$.
Equation (\ref{thicknesssolution})
implies that $h \rightarrow 2/\sigma$, $(2/m_c)^{1/2}x^{3/2}$,
and $2\sigma/[b_{r,s}^2+b_{\phi,s}^2]$ in the limits where
self-gravity, central-mass gravity, and magnetic stresses, respectively,
dominate the vertical squeezing of the disk.\footnote{CCK employed an
expression similar to equation (\ref{thicknessequation}), except
that they omitted the magnetic terms. Their solution for $h$
thus has the same form as equation (\ref{thicknesssolution}),
but with $\hat m_c$ and $\hat \sigma$ replaced by $m_c$ and
$\sigma$, respectively.}

As discussed in \S~\ref{formulation}, the initial ($t=0$)
conditions [which in the self-similar model also represent the outer
asymptotic ($r\rightarrow \infty$) values] correspond to a
collapsing core just before PMF. Thus we require
\begin{equation}
\label{initialconditions}
\sigma\rightarrow\frac{A}{x},
\ b_z\rightarrow\frac{\sigma}{\mu_0},
\ u\rightarrow u_0,\ v\rightarrow v_0\ \ {\text{as}}\ x\rightarrow \infty\, ,
\end{equation}
where $A$, $\mu_0$, $u_0$, and $v_0$ are constants.
As discussed in \S~\ref{formulation}, our fiducial values for
$A$ ($=3$) and $u_0$ ($=-1$) are the same as those used by
CCK; they are based on the results of numerical
simulations of nonrotating cores and are compatible with
observations.  We similarly adopt the CCK choice
$\mu_0=2.9$. Our reference values for the uniform initial
rotation $v_0$ are based on the range of measured core angular
velocities and typical cloud parameters that enter into the estimate (\ref{v0}).

\section{Results}
\label{results}
We have studied three distinct cases of rotational core
collapse, which we present in this section in order of increasing
complexity. They are: (1) A purely hydrodynamical collapse, with
no mechanism for angular momentum transport (\S~\ref{nonmagnetic});
(2) An ideal-MHD (IMHD) collapse, which incorporates magnetic braking but
does not include any magnetic field diffusivity that could
prevent the buildup of a central magnetic
monopole (\S~\ref{MHD}); (3) an MHD collapse that includes ambipolar diffusion
(AD; \S~\ref{AD}). The latter model is able to reproduce the basic
observational features of rotationally supported protostellar
disks and their central YSOs.

Our models have many features in common, arising from the
basic interplay between the centrifugal force and gravity.
Initially and at large distances from the center
(i.e., for large values of $x$),
gravity is stronger than the centrifugal force
and matter falls in at a high, supersonic
speed.  In this regime $m\gg m_c$, and
self-gravity dominates over attraction by the central object.
A very different behavior characterizes
small values of $x$. Typically, the innermost region constitutes
a Keplerian accretion disk, where
gravity approximately balances the centrifugal force and where the
infall speed is low and subsonic.
The transition between the supersonic and subsonic inflow regimes
is achieved through a shock, which is located at the point ($x_c$)
where the infalling matter encounters a centrifugal
barrier. Inward of this shock the column density increases
significantly, forming a dense accretion disk. The location of
the centrifugal shock roughly coincides with the {\em centrifugal
radius}, which is the largest radius where the
gravitational ($-m/x^2$) and centrifugal ($j^2/x^3$)
accelerations are in approximate balance
\citep[e.g.,][]{Basu1998}. In fact, we find that the point
where the equality $j^2(x)/m(x)=x$ is
first satisfied as the matter flows in yields a value that is
only a slight ($\sim 4\%$) overestimate of the actual shock location in
both the
IMHD and AD cases when (as is the case in our typical models)
$m(x_c)\approx m_c$. The value of
the centrifugal radius is sometimes estimated from the $t=0$
expressions for the surface density and angular momentum
profiles. This estimate could be rather inaccurate since it does not
account for the mass accumulation at the center or for the
loss of angular momentum that the infalling matter experiences even before it
reaches the centrifugal radius. We discuss more accurate
estimates of $x_c$ in the following subsections.

Our ``standard'' solutions, corresponding to the fiducial
parameter values, are presented in \S~\ref{MHDstandard} for the
IMHD case and in \S~\ref{ADstandard} for the AD case.
We also present solutions in the limit when the initial rotation
is very fast and magnetic braking
is either nonexistent (\S~\ref{nonmagnetic}) or comparatively
inefficient (\S~\ref{MHDfast} and \S~\ref{ADfast} for the IMHD
and AD cases, respectively). Under these conditions the
rotationally supported disk becomes fairly extended and its mass
greatly exceeds
that of the central object. Another ``extreme'' type of a solution
is obtained in the opposite limit, when the initial rotation is
braked so efficiently that $j$ is effectively reduced to zero at
a finite value ($x_j$) of $x$. In this limit, too, we present solutions
for both the IMHD (\S~\ref{MHDstrong}) and AD
(\S~\ref{ADstrong}) cases. These configurations are effectively
nonrotating (and thus resemble the CCK solutions) for $x<x_j$,
and they do not feature either a centrifugal
radius or a Keplerian accretion disk; instead, matter accretes
directly onto the central object at a supersonic speed.

\subsection{Nonmagnetic Rotational Collapse}
\label{nonmagnetic}
This problem was already treated in detail by \citet{SH}; it is
included here for reference and comparison, and to serve as a starting point.

In the absence of magnetic fields or of any other means of angular momentum
transport, the self-similar collapse is characterized by
$j/m=\Phi$, where $\Phi$ is a constant fixed by the
initial conditions ($\Phi = v_0/A$).
The inner asymptotic solution ($x\rightarrow 0$) is given by
$\sigma=E/x$ and $m=E x$, with $E\equiv[1+(1-4\Phi^2)^{1/2}]/(2\Phi^2)$.
Numerically, we proceed by applying the initial conditions at
$\xmax=10^2$ and integrating the differential equations
toward smaller values of $x$
until the expected location of $x_c$ (see eq.\ [\ref{x_c_nr}]
below) is reached. There we impose the isothermal shock jump
condition described in Appendix~\ref{isothermalshock},
and continue the integration until the asymptotic solution
is approximately fulfilled.
The exact value of $x_c$ is obtained iteratively by enforcing the matching to
this asymptotic solution to a high precision.

Figures \ref{1} and \ref{2} show profiles of mass, column density, and radial
speed for two kinds of initial rotation:
a typical ($v_0=0.1$) and a rapid ($v_0=1$) one.
In both cases the centrifugal shock is very abrupt, producing a large
increase in column density and a sudden decrease in inflow speed.
In the case with faster initial rotation, the effect of the
centrifugal barrier is so strong that it forces the radial
velocity $u$ to change sign below $x_c$ to $u>0$ (indicating a
backflow), as shown in Figure \ref{2}. The centrifugal shock thus separates
regions of
infall and backflow; it stays stationary in self-similar coordinates,
implying that it moves outward in physical space.
This phenomenon was already discussed in \citet{SH}.
By trying a few other values of $v_0$ (but keeping the other
parameter values unchanged), we have found that
$v_0=0.5$ is too small to produce a backflow but that $v_0=0.8$
is already large enough.

An interesting behavior exhibited by these solutions (and
standing out particularly in Fig.\ \ref{1}) is the development of a
mass ``plateau'' when the infall speed starts to increase above
its outer asymptotic ($x\gg 1$) value. This behavior is generic to all
except our fast-rotation solutions (\S~\ref{MHDfast} and
\S~\ref{ADfast}) and can be understood as follows. According to
equations (\ref{mw}) and (\ref{mdot}), $m=\sigma x(x-u)$
and $\dot m=-\sigma xu$. These equations are consistent with the
outer asymptotic solution $m=Ax=\sigma x^2$ and $u=u_0<0$
(eq.\ [\ref{initialconditions}]) only so long as $x\gg |u|$ (in
which case $m\gg \dot m$). This\columnbreak{}
\mbox{\plotone{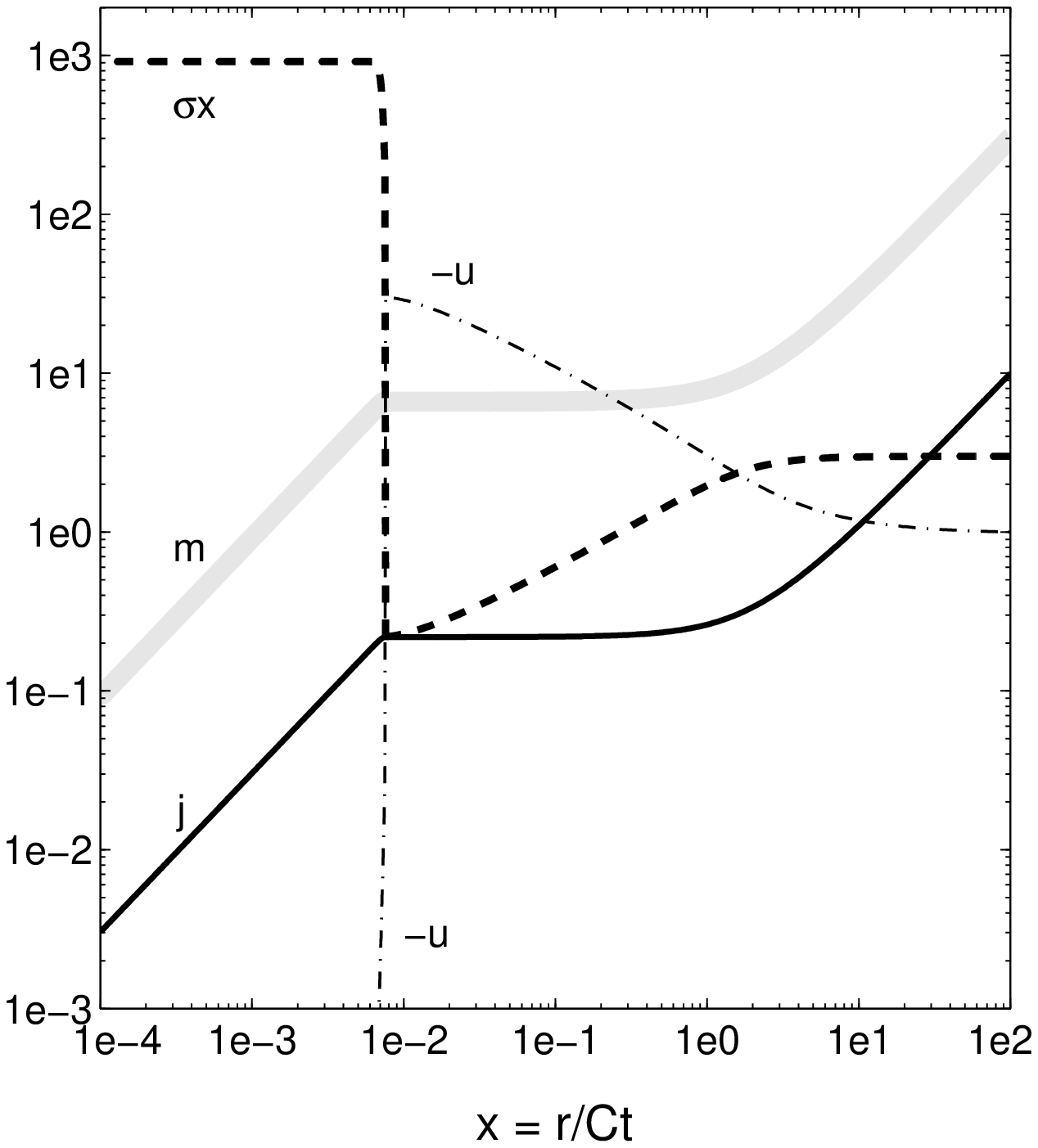}}

{\figcaption[f1.eps]{\label{1}Self-similar solution for a nonmagnetic
rotational
collapse. The variations of the normalized radial infall speed
$-u$, surface density $\sigma$, and total mass $m$ are
plotted as functions of the similarity variable $x$. The
initial (or outer asymptotic boundary) conditions correspond to
the parameter values $v_0=0.1$, $A=3$, and $u_0=-1$. The
centrifugal shock is located at $x_c=7.5{\times}10^{-3}$. No
central mass forms in this case on account of the assumed lack of any
angular-momentum transport mechanism.}}

\vskip 7ex

{\plotone{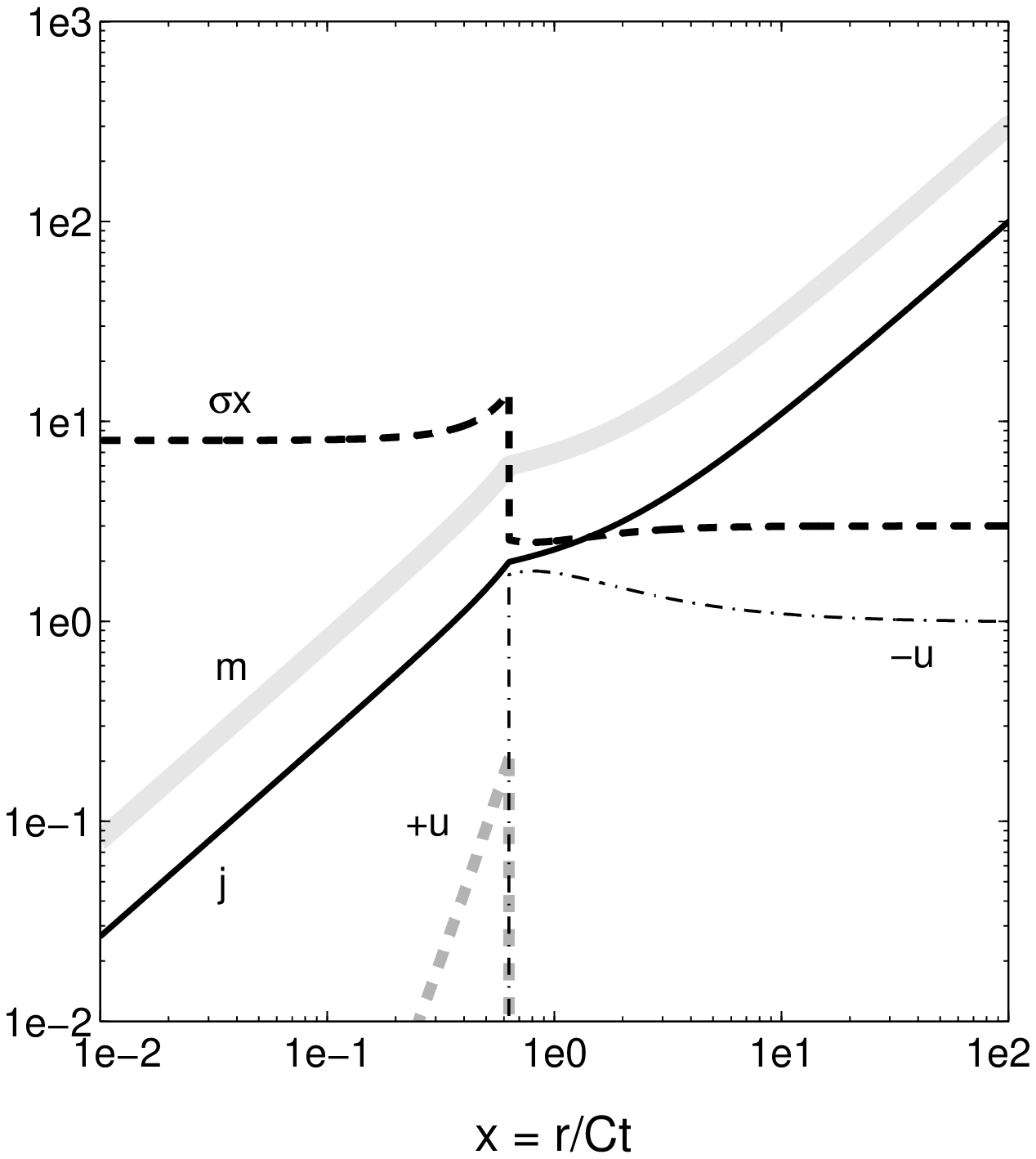}

\figcaption[f2.eps]{\label{2}Same as Fig.\ \ref{1},
except that the azimuthal-velocity
parameter $v_0$ is increased from 0.1 to 1.
Note the change in the sign of $u$ across the centrifugal shock in this case
and the comparatively large value of $x_c$ (=0.63).}}\columnbreak{}

{\noindent{}inequality breaks down when $x$
becomes small enough that it drops below $|u|$. This point
marks the approximate outer edge of the mass plateau, and we
label it $\xplateau$. Its value can be estimated by setting
\begin{equation}\label{x_pl}
\xplateau \approx |u_0|\, ,
\end{equation}
although typically $|u|$ is already somewhat larger than $|u_0|$
at that point. The plateau mass can then be approximated by
\begin{equation}\label{m_pl}
\mplateau \approx (A/\xplateau) \xplateau (\xplateau-u_0)\approx 2|u_0|A\, .
\end{equation}
For $x\ll \xplateau$,  $|u|\gg x$, and hence $m\approx\dot m$
and (using eq.\ [\ref{msigma}]) $dm/dx =\sigma  x \ll m/x$. The
latter inequality
accounts for the appearance of the plateau in the logarithmic plots of
$m$ vs.\ $x$. If $x_c/\xplateau$ is sufficiently small, then, in the
plateau regime, $|u|$ approaches the free-fall
value corresponding to $m \approx \mplateau$.
Hence, once $|u|$ comes to exceed $x$, their ratio
increases rapidly with decreasing $x$, and the mass plateau is quickly
established. The derivation of equations (\ref{x_pl}) and
(\ref{m_pl}) is independent of the
degree of magnetization of the flow, and
the expression (\ref{m_pl}) therefore also provides an excellent
first guess of $m_c$ in our fiducial
magnetic solutions that involve point-mass
formation.
These results do not, however, apply to the
fast-rotation cases considered in \S~\ref{MHDfast} and
\S~\ref{ADfast}, where a plateau does not form because the
centrifugal shock is established before $x$ becomes $\ll |u|$,
and where the central mass is typically smaller than that of the disk.}

Setting $j/m=\Phi$ and $m\approx \mplateau$ in the expression $x_c\approx
(j/m)^2 m$, we infer
\begin{equation}\label{x_c_nr}
x_c\approx 2 |u_0| v_0^2/A\, .
\end{equation}
We have used this expression to obtain the initial guess in our numerical
calculation. For the parameters of Figure \ref{1},
this gives $x_c \approx 7 \times
10^{-3}$, which is seen to be an excellent approximation to the
actual value. This estimate is only applicable if $x_c \lesssim
\xplateau$, i.e., for $v_0 \lesssim (A/2)^{1/2}$. This condition
is still marginally satisfied in the solution presented in
Figure \ref{2}.

The inner asymptotic solution of this model precludes the formation of a
point mass.  As the azimuthal speed $v$ cannot diverge
as $x\rightarrow 0$, $j=vx$ must vanish at the origin.
To form a central object with $m_c\neq 0$, a braking mechanism that
allows $j/m$ to decrease as $x\rightarrow 0$ is required.
The absence of such a mechanism explains the linear decline of
$m$ with $x$ at small values of $x$ (which for the parameters
used in Figs.\ \ref{1} and \ref{2} effectively starts already at $x_c$).
It may also be expected that a braking mechanism
will lead to a more gradual radial deceleration, reducing
or even altogether preventing the backflow exhibited by the rapid-rotation
solution shown in Figure \ref{2}. In the following subsections we
examine to what extent these expectations are met in the
presence of magnetic braking.

\subsection{Ideal-MHD Rotational Collapse}
\label{MHD}
In this section we present an IMHD model for magnetic braking
based on the formalism described by equations
(\ref{uequation})--(\ref{thicknessequation}), in which we set $\eta=0$.

The existence of a braking mechanism leads to a qualitative
change in the character of the solutions. In contrast to the
nonmagnetic solution presented in \S~\ref{nonmagnetic}, $j/m$ is
not a constant in this case, and magnetic
braking allows a central mass to form despite the
initial presence of rotation.

Flux freezing under the ideal-MHD assumption implies that
$b_z=\sigma/\mu$ and $\psi=m/\mu$, where $\mu$, the
nondimensional mass-to-flux ratio, is a constant (equal to its
initial value $\mu_0$). It is also convenient to introduce
the auxiliary constant $\mubar\equiv(1-\mu^{-2})^{1/4}$,
which is useful for describing the effect of the magnetic ``dilution'' of the
gravitational field \citep{ShuLiIsopedic}.

There are two qualitatively different sets of asymptotic relations
for the inner region ($x\rightarrow 0$) of this model.
The first set, presented in
equations (\ref{MHDasymptoticbegin})--(\ref{MHDasymptoticend})
below, applies to the configurations described in
\S~\ref{MHDstandard} and \S~\ref{MHDfast}.
The second set, presented in \S~\ref{MHDstrong}, applies only in
the strong-braking case. Both of these sets describe the
formation of a dynamically dominant central mass and a
split-monopole field. However, the first set represents a
rotationally supported disk, whereas the second one depicts a
nonrotating, supersonic inflow.

The disk-like inner asymptotic solution is
\begin{eqnarray}
\label{MHDasymptoticbegin}
\dot m= m&=&m_c\, ,\\
\label{jMHDasymptotic}
j&=&\mubar^2m_c^{1/2}x^{1/2}\, ,\\
-u=w&=&(2\delta)^{1/2}m_c^{1/4}\mu^{-1}\mubar^{-1}x^{1/4}\, ,\\
\label{bzMHDasymptotic}
-b_{\phi,s}/\delta=b_z=\sigma/\mu&=&\mubar
m_c^{3/4}(2\delta)^{-1/2} x^{-5/4}\, ,\\
\label{brMHDasymptotic}
b_{r,s}=\psi/x^2&=&m_c/(\mu x^2)\, ,\\
h=2\sigma/b_{r,s}^2&=&\mubar\mu^3m_c^{-5/4}(2/\delta)^{1/2}x^{11/4}\, .
\label{MHDasymptoticend}
\end{eqnarray}
These equations represent an essentially Keplerian disk; in
particular, the expression for $j$ differs from that of a purely
Keplerian system only by the magnetic-dilution factor $\mubar^2$
[which, however, remains close to 1
($\approx 0.94$) for our fiducial value of $\mu_0$].
The $j$ profile corresponding to such a disk implies that $j/m$
decreases with decreasing $x$. The magnetic braking required to account for
this change from the initial distribution
fixes the strength and radial power-law index of the magnetic
field components $b_z$ and $b_{\phi,s}$. The flux-freezing
condition then determines the value of $\sigma$, and the mass
conservation relation, in turn, yields the expression for $u$.

The behavior of the IMHD solutions is determined primarily
by the magnitude of the initial rotation and the strength of the
magnetic braking. We consider three qualitatively different cases:
a typical collapse in which both effects are relevant
over most scales (\S~\ref{MHDstandard});
a limiting case corresponding to
weak braking and fast rotation (\S~\ref{MHDfast}); and
another limiting case in which very strong braking is capable of
eliminating rotation altogether by the time the flow reaches
a finite distance from the origin, resulting in the formation of an
inner nonrotating-inflow zone (\S~\ref{MHDstrong}).

The method used to solve the differential equations is essentially the
same as that used in \S~\ref{nonmagnetic}. For the initial guess
of the value of $x_c$ we again use the estimate (\ref{x_c_nr}).
Although magnetic braking does operate in this case, the
deviation of $j/m$ from its $x\gg 1$ asymptotic
value remains relatively small in the range $x>x_c$, and the
approximation obtained for the nonmagnetic case remains fairly
good (as evidenced by the fact that the value of $x_c$ in the
solution presented in Fig.\ \ref{3} differs by a factor of less than 2 from its
value in Fig.\ \ref{1}). The discontinuities in the variables
$\sigma$, $b_z$, and $u$ across the shock are estimated using
the appropriate shock jump conditions. For the typical case
(\S~\ref{MHDstandard}) the shock occurs in a region where
$h\approx 2\sigma/b_{r,s}^2$ and the
``magnetically squeezed shock'' jump conditions described in
Appendix~\ref{magneticshock} are applicable.
In the fast-rotation case (\S~\ref{MHDfast}) the shock is located
sufficiently far away from the origin that a generalized (to
include magnetic effects) isothermal shock
(Appendix~\ref{generalizedisothermalshock}) provides a better approximation.
(The issue of shock jump conditions is not relevant to the strong-braking
inflow considered in \S~\ref{MHDstrong}, since no centrifugal shock
forms in that case.) As in \S~\ref{nonmagnetic}, a successful convergence
of the solution at $x<x_c$ to its asymptotic ($x\rightarrow 0$) form
requires a correct selection of the point $x_c$, which is achieved
iteratively. (Incorrect choices of $x_c$ produce
numerical solutions with a spontaneous singularity at some point
$0<x<x_c$, similar to those found in \citealt{LiSpontaneousSingularities}.)
As the value of the central mass is not known before the numerical
integrations start, we adopt $m_c\approx \mplateau$
(eq.\ [\ref{m_pl}]) as an initial guess that is refined
iteratively until convergence is attained.

\subsubsection{Fiducial Solution}
\label{MHDstandard}
The results for this case are presented in Figure \ref{3}.
For the fiducial parameters $v_0=0.1$, $\alpha=0.1$, and $\delta=1$,
rotation and braking are both moderate.
Magnetic braking is strong enough to prevent
shock formation in the region where self-gravity
dominates (as in \S~\ref{MHDfast}), but it is not so strong as to
prevent the shock from forming altogether (as in \S~\ref{MHDstrong}).
The centrifugal shock thus occurs well inside the
region where the central mass dominates the gravitational field,
with $m(x_c)\approx m_c$.
The disk that forms for $x<x_c$ is
thus essentially Keplerian and has an angular momentum profile
$j\propto x^{1/2}$. The large inclination of the
magnetic field lines to the disk surface indicated by the inner
asymptotic solution ($b_{r,s}/b_z \propto x^{-3/4}$ as $x\rightarrow
0$; see eqs.\ [\ref{bzMHDasymptotic}] and [\ref{brMHDasymptotic}]) suggests
that the disk would likely drive a centrifugal
wind from its surfaces, which would reduce
$m_c$. We neglect this effect in the present discussion, but we
consider it in \S~\ref{discussion} and Appendix~\ref{wind} in
relation to the ambipolar-diffusion fiducial solution.

The evolution of the inflow can be characterized by dividing the range
of $x$ into four regions, which are, in order of decreasing
distance from the center:

\paragraph{Self-gravity--dominated region.}
For the largest values of $x$,
the initial value of the ratio
$j/m$ ($=0.033$) is still conserved, and $m$, which is still much larger than
its central value, is proportional to $x$.

\paragraph{Free-fall region.}
After a transition region, where $j/m$ decreases slightly from its initial
value on account of magnetic braking, both $m$ and $j$
enter a plateau (at $x\approx \xplateau$; see eq.\ [\ref{x_pl}]). They become
independent of $x$, with values $\mplateau=6.2$ and
$\jplateau=0.18$ (corresponding to $j/m=0.028$). These computed values
can be substituted into the expression
$x_c\approx \jplateau^2/\mplateau$ to provide a refinement of the
estimate (\ref{x_c_nr}) for the location of the centrifugal
shock. As can be seen from a comparison of this value with the
computed result (listed in the caption to Fig.\ \ref{3}), the
approximation is excellent (as is the correspondence between the
value of $\mplateau$ and that of $m_c$).  Near the inner edge of
this region, the centrifugal force starts to exceed gravity.
\paragraph{The centrifugal shock.}
The outer edge of this region is where the variables
$\sigma$, $u$, and $b_z$ undergo the discontinuities prescribed
by the shock jump conditions of Appendix~\ref{magneticshock}.
Just inside the discontinuity the column density becomes very large, but it
then rapidly decreases toward its asymptotic near-Keplerian
value. The infall speed decreases rapidly, but not so strongly
as to produce any backflow (i.e., $u$ remains $<0$ for all $x$).
It is in this postshock transition zone
that the angular momentum makes the final
adjustment to its asymptotic, magnetically diluted,
Keplerian value: this sudden bout of braking is triggered by the spike
in $b_z$ that results from the shock compression.
The column density spike that occurs in this region is
associated with a small (but finite) mass, $\Delta m\approx 0.02$,
which\columnbreak{}
\mbox{\plotone{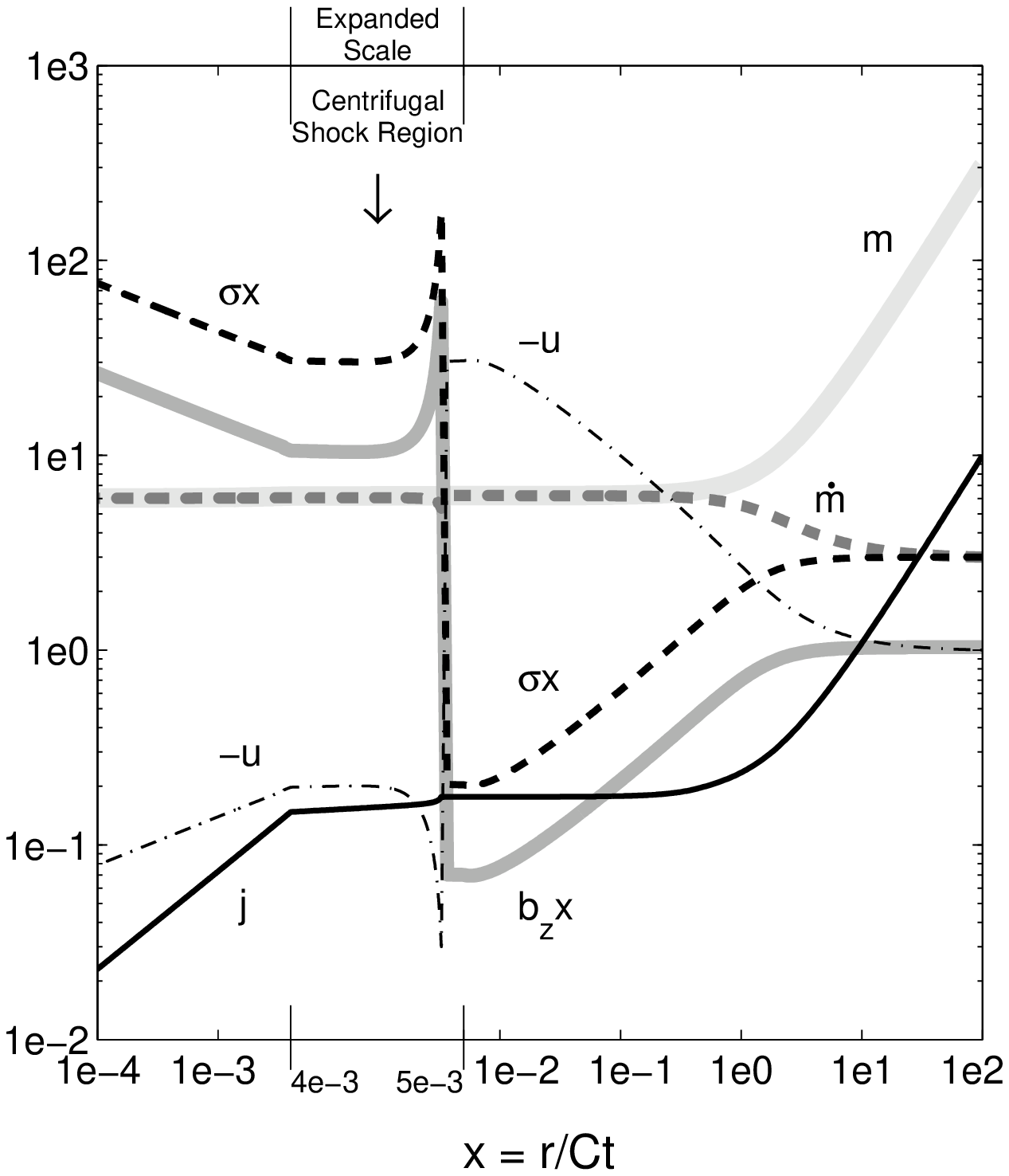}}

{\figcaption[f3.eps]{\label{3}Fiducial solution for an ideal-MHD rotational
collapse. The variations of the normalized radial infall speed
$-u$, surface density $\sigma$, total mass $m$, mass accretion
rate $\dot{m}$, specific angular momentum $j$, and $z$ component
of the magnetic field $b_z$ are
plotted as functions of the similarity variable $x$. The
initial (or outer asymptotic boundary) conditions correspond to
the parameter values $v_0=0.1$, $A=3$, $u_0=-1$, and
$\mu_0 = 2.9$, whereas the
magnetic-braking model parameters are $\alpha=0.1$ and $\delta=1$.
The centrifugal shock is located
at $x_c=4.9{\times}10^{-3}$, and the horizontal scale in its
vicinity has been expanded to show details of the
postshock transition zone. The central mass is $m_c=6.0$.}}

\vskip 5ex

{\noindent{}could be pictured as a massive ring located
at the edge of the Keplerian disk. After undergoing fast
variations as it transits through this region,
the flow merges seamlessly into the next, smoothly varying zone.}
\paragraph{Magnetically diluted Keplerian disk.}
In this region, the asymptotic equations
(\ref{MHDasymptoticbegin})--(\ref{MHDasymptoticend}) are
satisfied. The rotationally supported disk has a mass
$m_d\approx 3.2\%$ of $m_c$ and is in an almost perfect dynamical
equilibrium, as the inflow speed is low ($\lesssim 0.2\, C$).

\subsubsection{Fast Rotation}
\label{MHDfast}
If the initial azimuthal speed $v_0$ is high and the braking
parameters $\alpha$ and $\delta$ are moderate or small, then
the value of $j$ can become quite high, which can lead to
a large value of the centrifugal radius $x_c$.
In an extreme case, the centrifugal shock may occur so far
away from the center that it is located within the
self-gravity--dominated region, where $m\gg m_c$.

Figures \ref{4} and \ref{5} show
two solutions, corresponding to a fast initial rotation ($v_0 =
1.5$) and two different values of the braking parameter
$\alpha$ (0.01 and 0.1, respectively; $\delta =1$ in both cases).
The centrifugal shock in these solutions is so strong that the
radial speed $u$ changes sign across the shock.
This is similar to the backflow observed for large values of $j$
in the absence of braking (see Fig.\ \ref{2}) and is consistent with
the expectation that the braking must be strong enough to
qualitatively change the
behavior of a rapidly rotating inflow.
The backflow region in\columnbreak{}
\mbox{\plotone{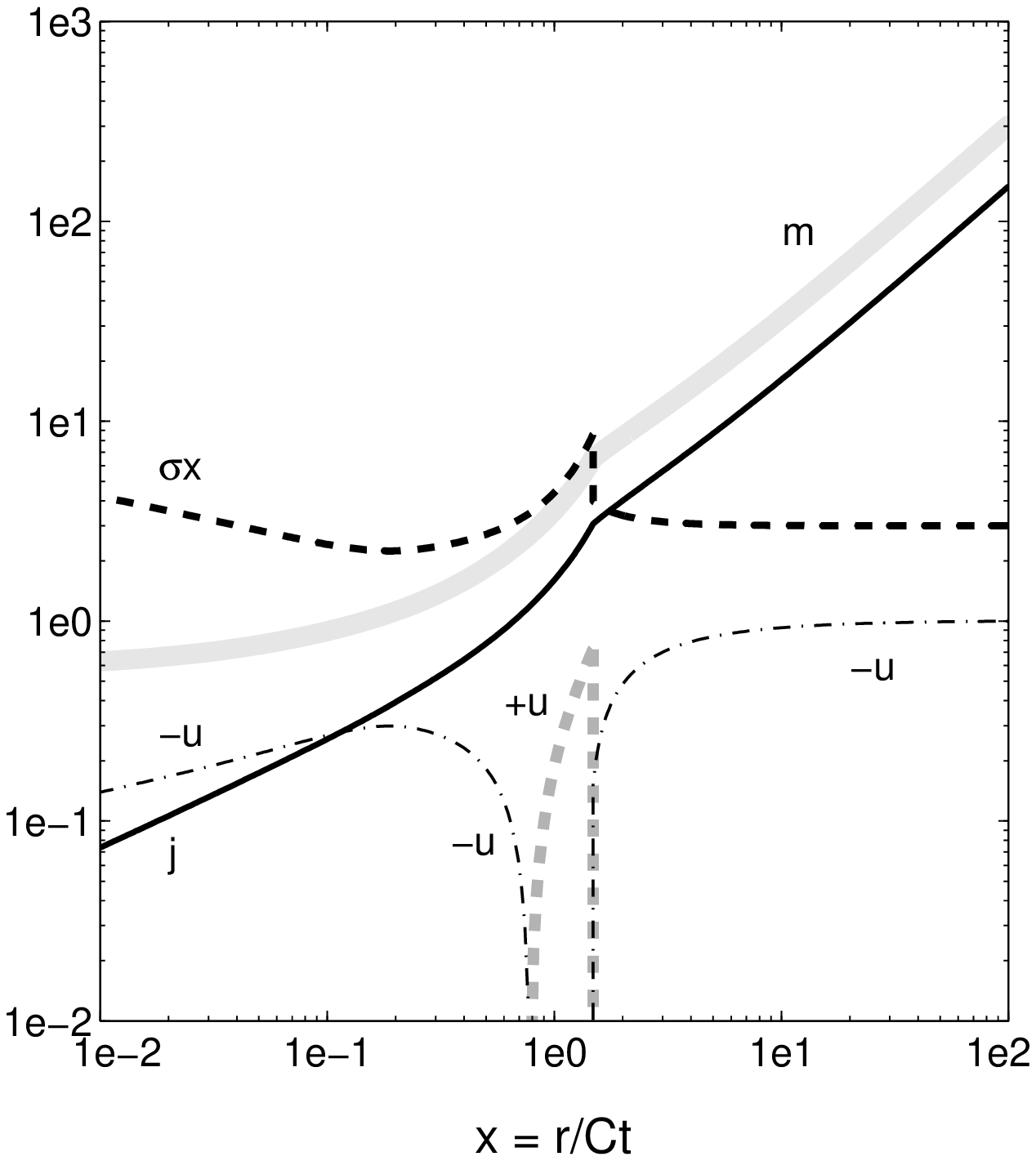}}

{\figcaption[f4.eps]{\label{4}Fast-rotation solution for an ideal-MHD
rotational
collapse. The variations of the normalized radial
infall speed $-u$, surface density
$\sigma$, total mass $m$, and specific angular momentum $j$ are
plotted as functions of the similarity variable $x$. The
model parameter values are the same as in the fiducial case
(Fig.\ \ref{3}), except that the azimuthal-velocity parameter $v_0$ is
increased from 0.1 to 1.5. In this case the centrifugal-shock
radius is comparatively large
($x_c=1.48$) and the central mass is rather small ($m_c=0.57$).
A backflow layer ($u>0$) is present just inside the centrifugal shock:
within this layer, the velocity curve depicts $+u$ instead of $-u$.}}

\vskip 4ex

{\plotone{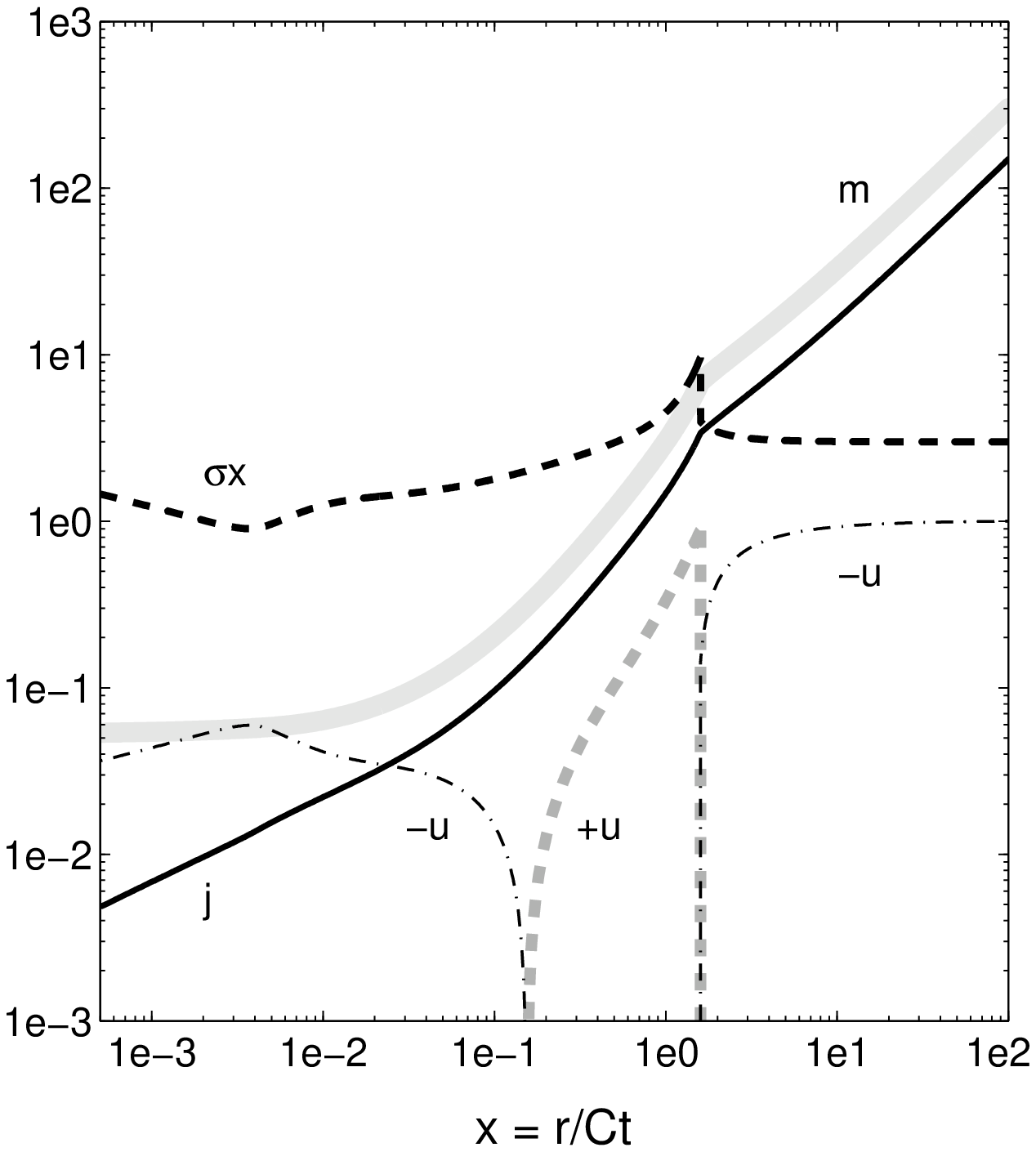}

\figcaption[f5.eps]{\label{5}Same as Fig.\ \ref{4}, except that the braking
parameter $\alpha$ is
decreased from 0.1 to 0.01. This results in a wider backflow
layer and a smaller Keplerian-rotation region. The
centrifugal shock location is not much changed ($x_c=1.53$), but
the central mass is strongly reduced ($m_c=0.05$).}}\columnbreak{}

{\noindent{}this case is clearly seen to occupy only a finite range
in $x$ and not to reach all the way to the origin: it
represents an annulus, interior to which the infall resumes.}

The large values of $x_c$ result in extended, and hence fairly
massive, disks. On the other hand, the comparatively weak
braking inhibits mass accumulation at the center: the central
masses obtained in these solutions (listed in the captions to
Figs.\ \ref{4} and \ref{5}) are much smaller than the mass derived for the
fiducial parameters in \S~\ref{MHDstandard} (which, in turn, is comparable to
$m_c$ in the nonrotating collapse model of CCK). The outer
regions of the disks are therefore manifestly non-Keplerian.
(The specific angular momentum approaches to within $\sim 20\%$
of the asymptotic, diluted-Keplerian value given by
eq.\ [\ref{jMHDasymptotic}] only when $x$ decreases below $\sim
0.11x_c$ for the solution in Fig.\ \ref{4} and below $\sim 0.04x_c$ for
the solution in Fig.\ \ref{5}.)
As can be seen from a comparison of Figures \ref{4} and \ref{5}, a larger value
of the braking parameter $\alpha$ leads to a reduction
in $j$ and results in the value of $x_c$ becoming smaller and
that of $m_c$ larger. Correspondingly, the width of the backflow
region is reduced and the Keplerian regime
extends further out. These trends continue as the effect of
rotation is diminished: by the time the parameters attain their
fiducial values, the backflow region has completely disappeared
and the inner asymptotic solution starts very close to $x_c$.

In summary, the collapse of fast-rotating cores that lack an
exceptionally strong braking mechanism results in
fairly massive disks and in comparatively low-mass central
objects. The disks are rotationally supported and largely
self-gravitating, with only their innermost regions exhibiting
Keplerian behavior.

\subsubsection{Strong Braking}
\label{MHDstrong}
If braking is very efficient then the innermost region of the
inflow can have effectively zero angular momentum even if the
collapsing core starts out rotating. This situation is
illustrated in Figure \ref{6}, which depicts a solution with a fairly
respectable initial rotation ($v_0=1$) but with extreme values
of the braking parameters ($\alpha=\delta=10$), which give rise
to a strong surface azimuthal field (see
eq.\ [\ref{bphiequation}]). In this case the angular momentum is
reduced effectively to zero at $x_j\approx 0.2$, before a centrifugal shock
could occur, and thus no rotationally supported disk is
formed. For $x<x_j$, $j=b_{\phi,s}=0$, and as
$x\rightarrow 0$, the inflow tends to the asymptotic form of
the nonrotating magnetic solution derived by CCK (see their
{\S}3.3). In this limit the flow becomes
a supersonic, magnetically diluted free fall onto the central mass:
\begin{eqnarray}
\label{MHDstrongasymptoticbegin}
\dot m = m&=&m_c\, ,\\
-u&=&[2m_c(1-\mu^{-2})/x]^{1/2}\, ,\\
\sigma=\mu b_z&=&[m_c/2(1-\mu^{-2})x]^{1/2}\, ,\\
b_{r,s}=\psi / x^2&=&m_c/(\mu x^2)\, ,\\
h=2\sigma/b_{r,s}^2&=&\mu^2[2/m_c^3(1-\mu^{-2})]^{1/2}\,x^{7/2}\, .
\label{MHDstrongasymptoticend}
\end{eqnarray}
The numerical solution converges to these asymptotic values
already quite close to $x_j$ (at $x\sim 0.1$).

Equation (\ref{MHDstrongasymptoticend}) for the disk scale
height shows the strong effect of magnetic squeezing by the
radial magnetic field component in the
asymptotic solution. (Since the radial surface magnetic field
scales as $x^{-2}$, it dominates the magnetic
terms in both the vertical and the radial force-balance
equations.) The dependence of $h$ on $x$ ($\propto x^{7/2}$) is
much stronger than in the solution presented by CCK ($h\propto x^{3/2}$),
in which the effect of magnetic pinching was neglected. However,
apart from this difference, these two solutions are very similar;
in particular, the value of\columnbreak{}
\mbox{\plotone{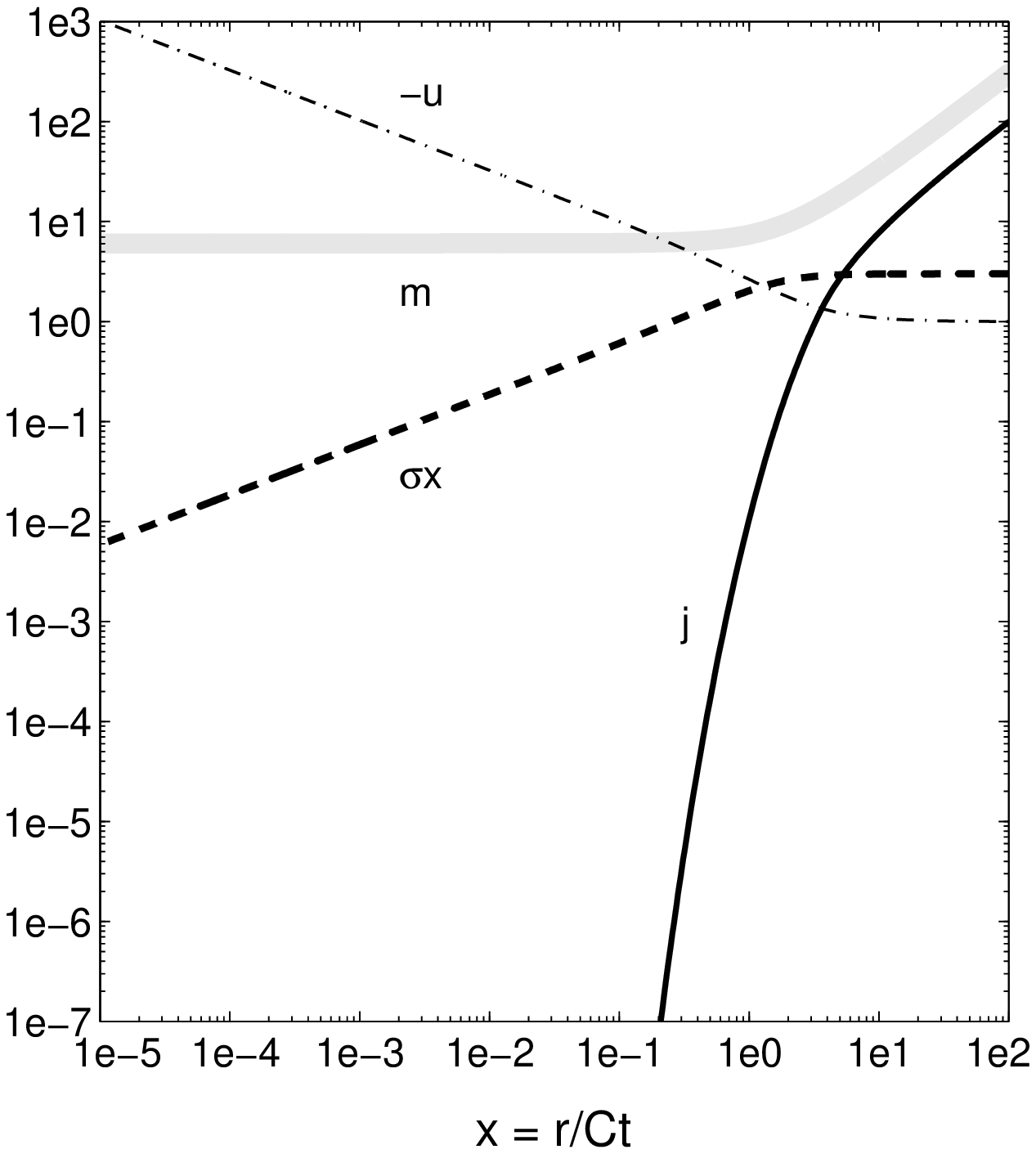}}

{\figcaption[f6.eps]{\label{6}Strong-braking solution for an
ideal-MHD rotational
collapse. The variations of the normalized radial
infall speed $-u$, surface density $\sigma$, total mass $m$, and
specific angular momentum $j$ are
plotted as functions of the similarity variable $x$. Moderately fast
initial rotation ($v_0=1$) and large braking parameters
($\alpha=\delta=10$) are assumed; the other parameter values are
the same as in the fiducial solution of Fig.\ \ref{3}. In this case
$j$ decreases to a very small value at a finite distance from the
origin ($x_j\approx 0.2$), and no centrifugal shock is
established. The central mass is $m_c=6.06$.}}

\vskip 5ex

{\noindent{}the central mass quoted by CCK
($m_c=6.1$) is very close to the value that we derive ($m_c=6.06$).
We note, however, that the extremely rapid decrease of $h$ with
radius that is implied by this solution is unlikely to be sustained in
reality. For example, when the disk becomes sufficiently thin,
the radial field components on opposite sides of the disk might
reconnect, resulting in an upper bound on
$b_{r,s}$. Furthermore, the decrease in $h$ will be
arrested once the column density grows to a value for which the
disk becomes opaque and the isothermal approximation breaks
down. We do not consider these caveats in more detail since (as
we point out at the beginning of \S~\ref{AD}) we expect {\em radial}
field diffusion
to intervene (possibly even before the above effects come into
play) and prevent the IMHD configuration from being set up near the center.}

\subsection{Ambipolar Diffusion-Dominated Rotational Collapse}
\label{AD}
Using the $x\rightarrow 0$ asymptotic relations for the IMHD solution given by
equations (\ref{MHDasymptoticbegin})--(\ref{MHDasymptoticend}),
it can be readily verified that the nondiffusive terms in the induction
equation (\ref{psiequation}) are $\propto x^{0}$ whereas the diffusive
term (associated with the magnetic tension force) is $\propto
x^{-1/4}$, so that the latter will become
dominant near the origin. It follows that the IMHD solution
cannot apply in the vicinity of the center in cases in which an
accretion disk is predicted to form. Ambipolar diffusion must
therefore be incorporated into the model, and in this subsection
we describe how the nature of the solutions changes when this is
done.

In the case of the asymptotic solution for a
strongly braked rotational infall or a nonrotational collapse
(eqs.\ [\ref{MHDstrongasymptoticbegin}]--[\ref{MHDstrongasymptoticend}]),
the nondiffusive terms in equation (\ref{psiequation}) are still
$\propto x^{0}$, but the diffusive term scales as $x^{1/2}$ and remains
subdominant as $x\rightarrow 0$. The asymptotic IMHD solution is thus
self-consistent in this case, as we verified by checking
that, in the model presented in Figure \ref{6},
the diffusive term is everywhere
smaller than 10\% of the flux-advection term (assuming $\eta\lesssim
1$). This conclusion differs from that of CCK, a discrepancy
that can be traced to their neglect of the magnetic squeezing
term in the equation for $h$. However, as we noted in \S~\ref{MHDstrong},
real disks are unlikely to become as extremely thin as
the IMHD solution that incorporates this term would imply. Furthermore,
the accumulation of magnetic flux at the
center, which is indicated by the IMHD split-monopole solution, would
lead to a severe magnetic flux problem (see
\S~\ref{introduction}) if allowed to proceed and is likely
prevented from occurring by the action of {\em Ohmic}
diffusivity in the disk or the YSO \citep[see][]{LiMcKee1996}.

The ascendancy of the ambipolar diffusion term at low values of
$x$ means that ions and neutrals behave differently in that
region. In contrast, the initial ($x\gg 1$) conditions
involve no field diffusivity and correspond to negligible drift
between these two particle species. The study of these two
regimes can be facilitated by observing that, in most cases,
the strong inequality $b_{r,s}\gg h (db_z/dx)$ is satisfied
for almost all values of $x$: the gradient of $b_z$
is never very large (except in shocks), and the disk~is usually very thin.
Under these conditions, one can treat the flux conservation relation
(eq.\ [\ref{psiequation}]) as a quadratic equation for $b_z$,
\begin{equation}
\label{bzquadraticequation}
\eta \psi x^{-1}h^{1/2}\sigma^{-3/2}b_z^2
-x(x-u)b_z
+\psi=0\ ,
\end{equation}
where we regard the variables $\psi$, $h$, $\sigma$, and $u$ as
given. Equation (\ref{bzquadraticequation}) has two real roots,
$\blo$ and $\bhi$, and we have found that in most of the
solutions that we constructed they are well separated ($\blo \ll \bhi$).
This property makes it possible to obtain simple expressions for
$b_z$ in the IMHD~(large~$x$) and AD~(small~$x$) regimes.
Specifically, the lower root, which is applicable in the IMHD
regime, can be evaluated by disregarding
the quadratic term in equation (\ref{bzquadraticequation}):
\begin{equation}\label{blo}
\blo\approx \psi/xw=\sigma\psi/m\, .
\end{equation}
The vertical field component in the AD regime corresponds to the larger
root and
can be approximated by omitting the constant
term in equation (\ref{bzquadraticequation}):
\begin{equation}\label{bhi}
\bhi\approx \eta^{-1} x (\sigma/h)^{1/2} (m/\psi)\, .
\end{equation}

The transition between the IMHD and AD regimes can take place
either gradually or abruptly. The latter case occurs when the
flow is super--fast-magnetosonic and the flow variables are
changed by passing through a shock. The occurrence of such an
ambipolar diffusion-mediated shock was originally suggested by
\citet{LiMcKee1996} and subsequently verified in the nonrotating collapse
calculations of CK and CCK, which succeeded in resolving the
shock structure. We generalize these results to the
rotating-collapse case in \S~\ref{ADstandard}.
The most striking feature of this shock is the abrupt jump in
$b_z$ between the two roots (\ref{blo}) and (\ref{bhi})
(the other flow variables are also affected, but less strongly).
The jump represents a large local gradient in $b_z$, which
implies that the term $h (db_z/dx)$ in equation
(\ref{psiequation}) can no longer be disregarded. In fact, it is
this term that mediates the transition between the two roots of
equation (\ref{bzquadraticequation}) and allows the shock
structure to be resolved. The transition, however, remains very
sharp, and equation (\ref{bzquadraticequation}) remains
applicable outside a narrow neighborhood of the shock position $x_a$.

As in the IMHD solution presented in \S~\ref{MHD}, we find it convenient
to distinguish
between the fiducial case (\S~\ref{ADstandard}) and the limiting
cases of fast rotation (\S~\ref{ADfast}) and strong braking
(\S~\ref{ADstrong}). In the fiducial case the braking is strong
enough to place the centrifugal shock well
within the AD shock ($x_c \ll x_a$).
The centrifugal shock again disappears in the
strong-braking case, but an AD shock is still present, as in
the nonrotating, diffusive solution obtained by CCK. However, in
the fast-rotation case the centrifugal shock is established when the
flow is still in the IMHD regime, and hence for $x<x_c$ the inflow
speed is low. In this case the transition between the IMHD and
AD regimes occurs smoothly instead of in a shock (with the two
roots of eq.\ [\ref{bzquadraticequation}] remaining close to each
other), although it is still mediated by the $h (db_z/dx)$ term
in the induction equation.

To solve the self-similar equations as a boundary-value problem,
it is necessary to know the asymptotic behavior for
$x\rightarrow 0$. For a flow that continues to rotate all the
way down to the origin (as in the fiducial and
fast-rotation solutions), the behavior in this limit is given by
\begin{eqnarray}
\label{asymptoticm}
\dot m=m&=&m_c\ ,\\
j&=&m_c^{1/2}x^{1/2}\ ,\\
\label{ADu}
-u=w&=&(m_c/\sigma_1)x^{1/2}\ ,\\
\label{ADsigma}
\sigma&=&
\frac{(2\eta/3\delta)(2m_c)^{1/2}}{[1+(2\eta/3\delta)^{-2}]^{1/2}}
x^{-3/2}\nonumber\\
&\equiv& \sigma_1x^{-3/2}\ ,\\
\label{ADbz}
b_z=-b_{\phi,s}/\delta&=&
[{m_c}^{3/4}/(2\delta)^{1/2}]x^{-5/4}\ ,\\
\label{ADpsistandard}
b_{r,s}=\psi/x^2&=&(4/3)b_z\ ,\\
\label{asymptotich}
h&=&\{2/[1+(2\eta/3\delta)^2]m_c\}^{1/2}x^{3/2}\ .
\end{eqnarray}
This solution represents a Keplerian disk in which all the field
components have comparable magnitudes. In contrast with the
corresponding IMHD solution with its split-monopole field configuration (see
eqs.\ [\ref{MHDasymptoticbegin}]--[\ref{MHDasymptoticend}]), in
this case the field--matter decoupling brought about by the ambipolar
diffusion results in a field that is not strong enough to either
dilute the Keplerian rotation or to dominate the vertical
compression. [The vertical squeezing of the disk in the
asymptotic AD solution is dominated by the tidal and self-gravity forces,
which contribute in the proportion $1\, :\, (2\eta/3\delta)^2$.]
The asymptotic solution in the strong-braking limit, which (as
in the corresponding IMHD case) involves a flow that is nonrotating near
the origin, is given in \S~\ref{ADstrong}.

The solutions presented in this subsection have been obtained by using
the following numerical procedure:
\begin{enumerate}
\item Solve a simplified system of equations by disregarding the
$db_z/dx$ terms in equations (\ref{forceequation}),
(\ref{psiequation}), and (\ref{thicknessequation}) to obtain
approximate values for the five variables
$\sigma$, $u$, $\psi$, $j$, and $b_z$ at a selected matching
point $x_m$. We have found that, to assure numerical stability,
it is best to choose this matching point
in the region where the effects of ambipolar diffusion
are just starting to become dominant.
\item
Integrate the full system of equations
both outward and inward from the matching point.
In the outward direction, the solution reaches out to the point
$\xmax=10^2$, where it must match the
prescribed initial conditions of the flow (given
by the parameters $A$, $u_0$, $\mu_0$, and $v_0$).
In the inward direction, the flow must match the
$x\rightarrow 0$ asymptotic solution (given by
eqs.\ [\ref{asymptoticm}]--[\ref{asymptotich}] in the fiducial
and fast-rotation cases, and by
eqs.\ [\ref{ADstrongasymptoticbegin}]--[\ref{ADstrongasymptoticend}]
in the strong-braking case), which also determines the value of $m_c$.
Except in the strong-braking limit, the flow must also incorporate a
centrifugal shock (which is located inside the
matching point for the fiducial solution and outside it in the
fast-rotation case). The discontinuities that some of the flow
variables experience at $x_c$ are handled using the appropriate
shock jump conditions (Appendix~\ref{singular}).
\item
Iterate on the values of the five flow variables at the matching point $x_m$
as well as on those of the central mass ($m_c$) and (if
applicable) of the centrifugal-shock position ($x_c$) until convergence
is reached.
\end{enumerate}

\subsubsection{Fiducial Solution}
\label{ADstandard}
Our ``standard'' solution, corresponding to $\eta=1$,
$v_0=0.73$, $\alpha = 0.08$, and $\delta=1$, is presented in Figure
\ref{7}. In this case the initial
rotation is not very fast and the braking is moderate (see
\S~\ref{MHDstandard}), so the ambipolar-diffusion shock
is located further away from the center than the centrifugal
shock ($x_a/x_c \approx 30$). To more clearly separate the
different flow regimes, we also show a solution where $v_0$ is
reduced to 0.18 (Fig.\ \ref{8}a), which has the effect of reducing $x_c$ by a
factor $\sim 10$, and another one (Fig.\ \ref{8}b) where, in addition,
$\eta$ is reduced
to 0.7, which has the effect of further increasing $x_a/x_c$ (to $\sim
1\times 10^3$).
We now summarize the distinguishing properties of these flow regimes.

\paragraph{Outer region ($x>x_a$): ideal-MHD infall.}
This region extends from
$\xmax=10^2$, where the initial conditions are applied,
to $x=x_a$, the location of the resolved ambipolar diffusion shock.
This region does not differ substantially from the IMHD case,
with the effects of ambipolar diffusion remaining minimal.  The approximation
$b_z \approx \blo$ (eq.\ [\ref{blo}]) is excellent, and it is
also quite accurate to set $\mu\equiv \sigma/b_z \approx
\text{const}$.
The outer asymptotic solution remains roughly applicable for $x\gtrsim 10$.
Although the centrifugal force is still dynamically unimportant, there is some
magnetic braking, especially for $x\lesssim 1$, and so the ratio
$j/m$ is not exactly constant. Over most of this region the infall speed is
governed by self-gravity ($m\gg m_c$), but closer to the inner edge
($x\sim 2$), infall starts to be dominated by the central mass.

\paragraph{The ambipolar diffusion shock ($x\approx x_a$).}
This shock marks the end of the IMHD regime.
It can be resolved as a continuous transition, although it may
contain a viscous subshock.
Its most notable feature is the rapid increase in $b_z$,
which grows from essentially $\blo$ at the outer edge of the
shock transition to $\bhi$ (eq.\ [\ref{bhi}]) at the inner edge.
Although the inequality $h\ll x$ continues to hold, the $db_z/dx$
terms in the structure equations are important in this
range.
[It is worth noting in this connection that the sharp spike exhibited by the
disk half-thickness
curve at the location of the shock is a consequence of the large
$b_z$-gradient term in eq.\ (\ref{thicknessequation}). The
extreme narrowness of the spike ($\Delta x \ll |u| h$)
indicates, however, that the
assumption of vertical hydrostatic equilibrium is not justified
at that location. In reality, the enhanced magnetic squeezing at
the shock will reduce the disk thickness to only a fraction of
the amount indicated by the equilibrium curve over the fluid
transit time through the shock. This apparent overshoot is,
however, a highly localized phenomenon: the solution obtained by
omitting the $db_z/dx$ term in eq.\ (\ref{thicknessequation})
has a very similar global structure.]
Immediately behind the shock the inward acceleration is
temporarily reversed, resulting in a thin overshoot layer where the flow
undergoes a weak outward acceleration.
The shock structure is basically the same as that found in CCK
for a nonrotational flow: for our fiducial parameters, rotation
is not yet dynamically significant at this location.

To obtain an estimate of the value of $x_a$ (which can be used
as an initial guess in the numerical solution), we make use of
the fact that, in the infall region that is located just
downstream from $x_a$, $b_{r,s}\approx b_z\approx \bhi$.
These approximate equalities imply $x^3\approx
(\eta\psi^2/m)(h/\sigma)^{1/2}$.
We evaluate this expression at $x_a$ by using the fact that,
for $x>x_a$, ideal MHD is approximately valid and hence
$m/\psi$ is not very different from its
initial value $\mu_0$, and that at this location $m$ is
already close to $m_c$ so that
$\psi\approx m_c/\mu_0 = \text{const}$.
For the solutions shown in Figures \ref{7}
and \ref{8}, the 
vertical squeezing at $x_a$ is controlled by magnetic pinching, so
$h\approx{}2x^4\sigma/\psi^2$ (see
eq.\ [\ref{thicknesssolution}]). These approximations yield an
estimate of the shock location,
\begin{equation}\label{x_a}
x_a\approx \sqrt{2}\eta/\mu_0\, ,
\end{equation}
which depends only on prespecified flow
parameters. Analogous (albeit less simple) algebraic
expressions can be written
down for the cases in which the 
vertical squeezing is dominated by the central tidal force or by self-gravity.

Since the magnetic flux contained within $x_a$ is essentially the same
as the flux that initially threaded the enclosed mass, it also nearly
equals the flux trapped in the central split monopole of the IMHD
solution. Ambipolar diffusion does not destroy the accumulated flux
but rather redistributes it between the origin and the shock
location. As the value of $x_a$ generally increases with the
diffusivity parameter $\eta$, the approximate flux conservation within
$x_a$ implies that the vertical field component ($\approx \bhi$)
behind the ambipolar-diffusion shock decreases with increasing $\eta$
(in particular, it scales as $\eta^{-2}$ when $x_a$ is given by
eq.\ [\ref{x_a}]).

A comparison of Figures \ref{8}a and \ref{8}b reveals that the
deceleration of the flow in the AD shock is stronger the lower
the value of $\eta$. This can be attributed to the inferred
dependence of $\bhi$ on $\eta$, which results in higher
magnetic tension and pressure forces downstream of the shock
when $\eta$ is reduced.
If $\eta$ becomes smaller than $\sim{}0.2$
(with the other parameters remaining unchanged), then
the localized decrease of the infall speed inward of $x_a$
becomes large enough to reduce $w\equiv x-u$ below 1.
But $w=1$ is the singular line of the AD system of equations,
corresponding to the critical (sonic) speed (see
Appendix~\ref{singular}). A crossing
of this curve signals a discontinuity in the flow parameters,
which in this case implies the presence of an unresolved viscous
subshock at the downstream end of the resolved MHD shock transition
\citep[see][]{LiSpontaneousSingularities}.

\paragraph{AD-dominated infall ($x_c < x < x_a$).}
Over most of the region between the two shocks the centrifugal force
remains dynamically unimportant, so matter moves almost in free
fall, with the gravitational force dominated by the
central mass. The ions and field are only weakly coupled to
the neutrals, with the ion radial speed being effectively zero
(except in the vicinity of $x_c$).
The vertical field component continues to be given
by $b_z\approx \bhi$ to a very good approximation.
In many respects this zone resembles the innermost region of
the nonrotating flow studied by CCK, especially when it is
fairly extended ($x_a\gg x_c$), as it is in the two solutions
presented in Figure \ref{8}. In particular, $b_z\propto x^{-1}$,
$\psi\approx x^2b_z$,
and (over a somewhat narrower range) $\sigma\propto w \propto x^{-1/2}$.
If this region is wide enough, magnetic braking
can reduce $j$ to an essentially constant
value, $j\approx\jplateau$. The mass is also well approximated
by its plateau value (see eq.\ [\ref{m_pl}]),
$m\approx\mplateau\approx m_c$.\columnbreak{}
\mbox{\plotone{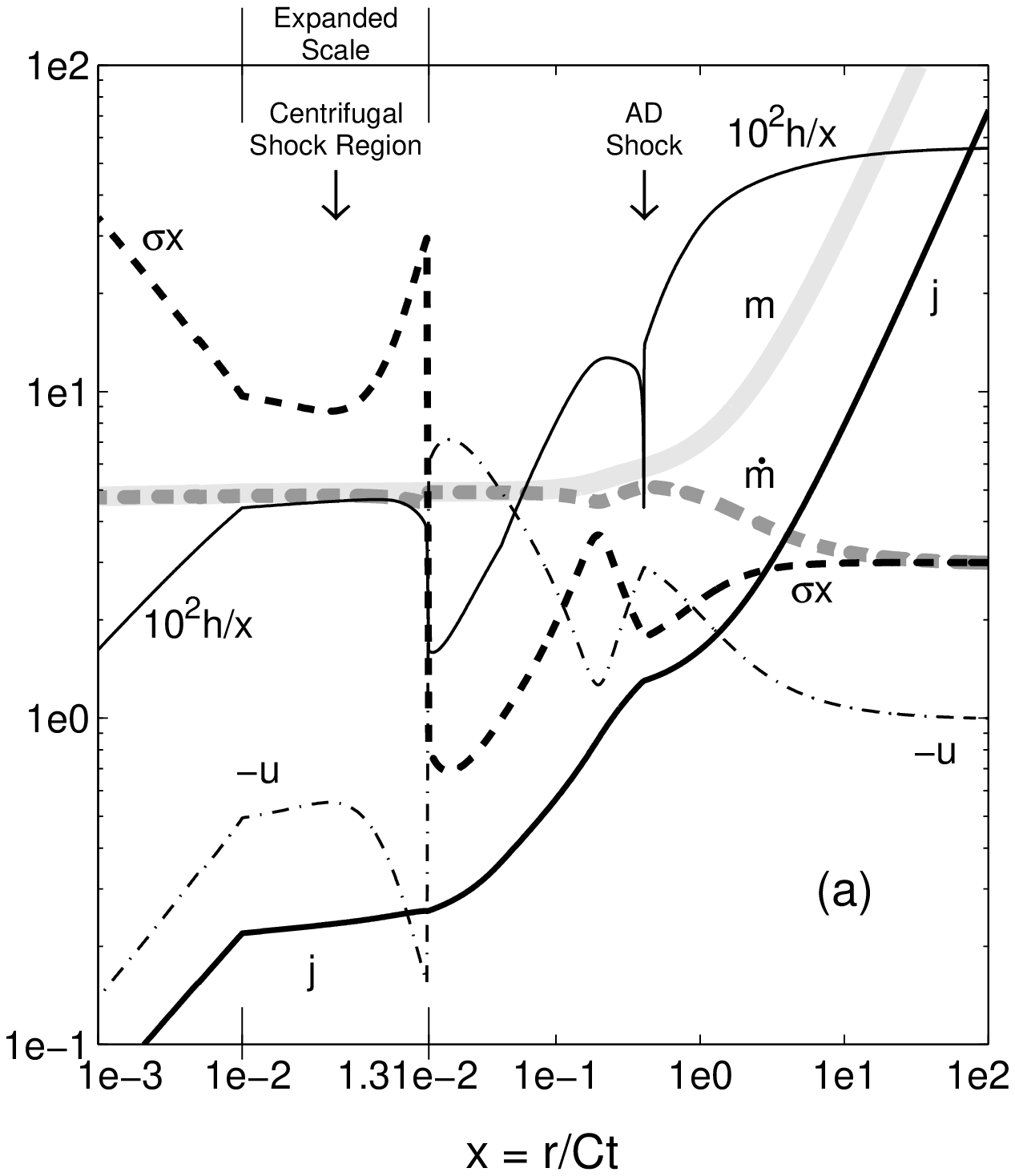}}

{\plotone{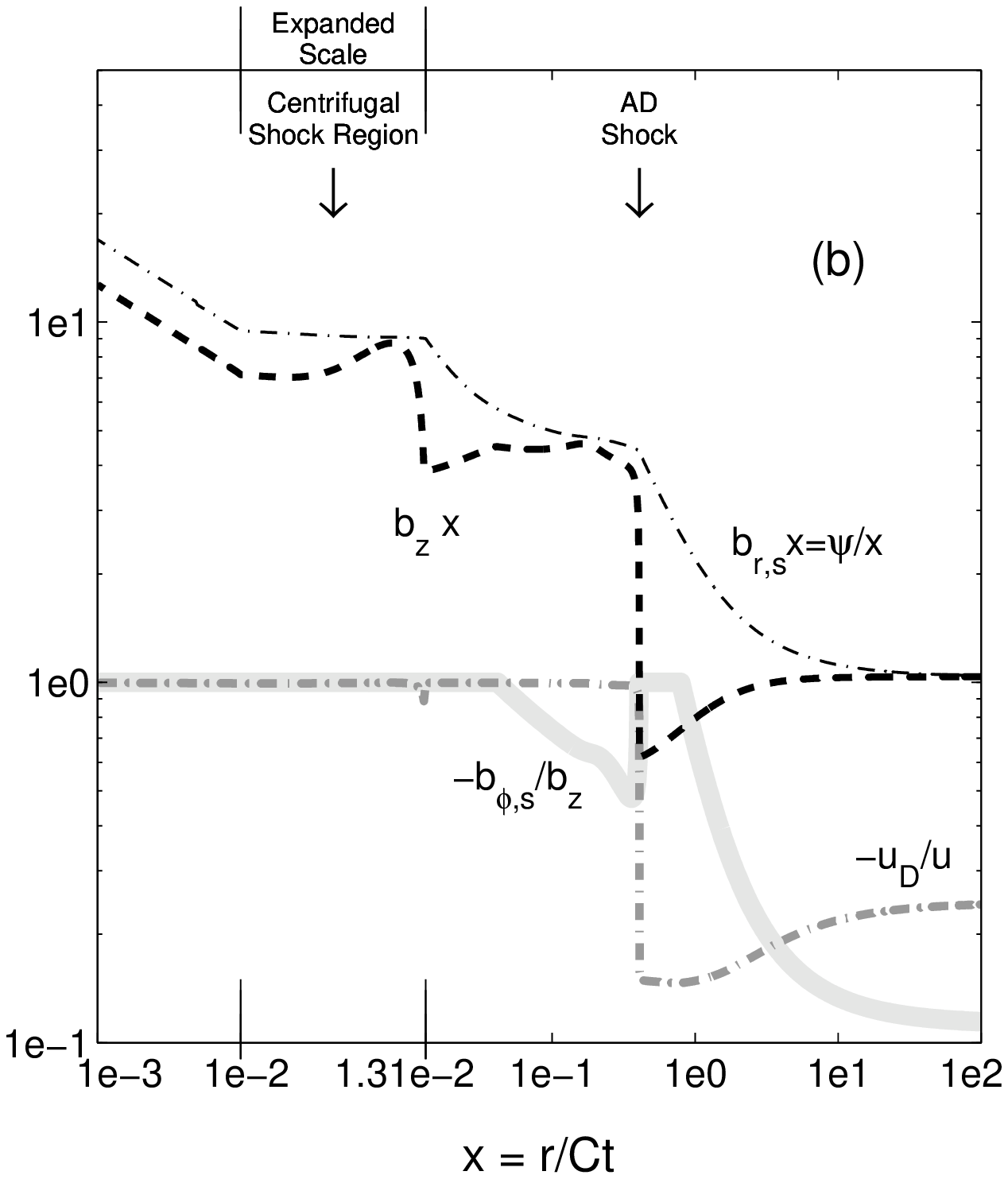}
\figcaption[f7a.eps,f7b.eps]{\label{7}Fiducial solution
for an ambipolar-diffusion rotational
collapse. The variations of the normalized radial infall speed
$-u$, surface density $\sigma$, total mass $m$, mass accretion
rate $\dot{m}$, specific angular momentum $j$, and the disk
half-thickness $h$ as functions of the similarity variable $x$
are plotted in Fig.\ \ref{7}a, and the corresponding variations of
the $z$ component
($b_z$) and the surface radial ($b_{r,s} = \psi/x^2$) and azimuthal
($b_{\phi,s}$) components of the magnetic field, as well as of
the normalized radial component of the
ion--neutral drift velocity ($u_D\equiv V_{D,r}/C$), are shown in
Fig.\ \ref{7}b.
The diffusivity parameter is $\eta=1$. The other parameter
values are $v_0=0.73$, $A=3$, $u_0=-1$, $\mu_0 = 2.9$,
$\alpha=0.08$, and $\delta=1$. The ambipolar-diffusion and
centrifugal shocks are located at $x_a=0.41$ and
$x_c=1.3\times{}10^{-2}$, respectively. The
central mass is $m_c=4.7$, and the plateau values in the
AD-dominated infall region are $\mplateau\approx 4.9$ and
$\jplateau\approx 0.26$.}}

\vskip 2ex

{\plotone{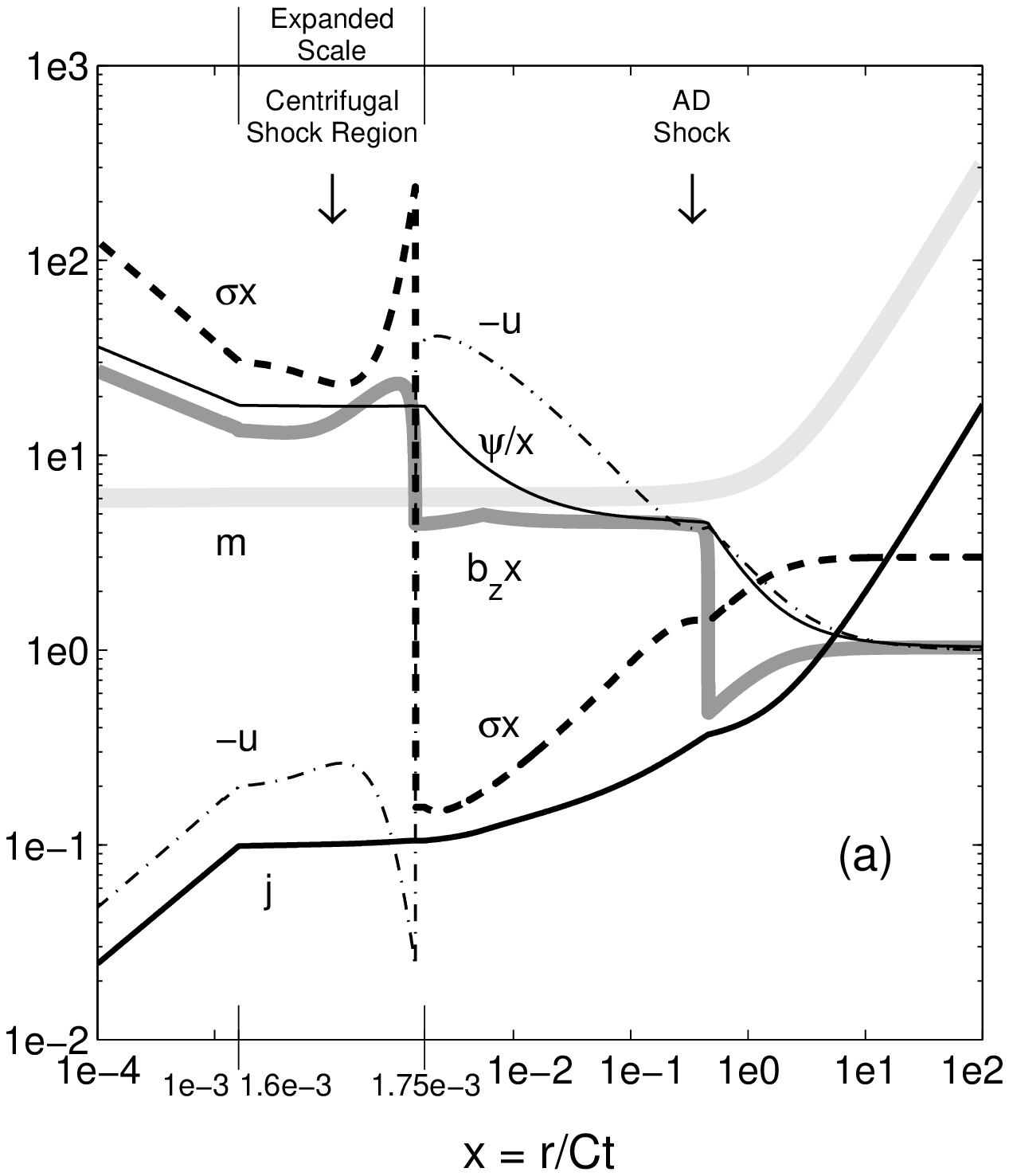}

\plotone{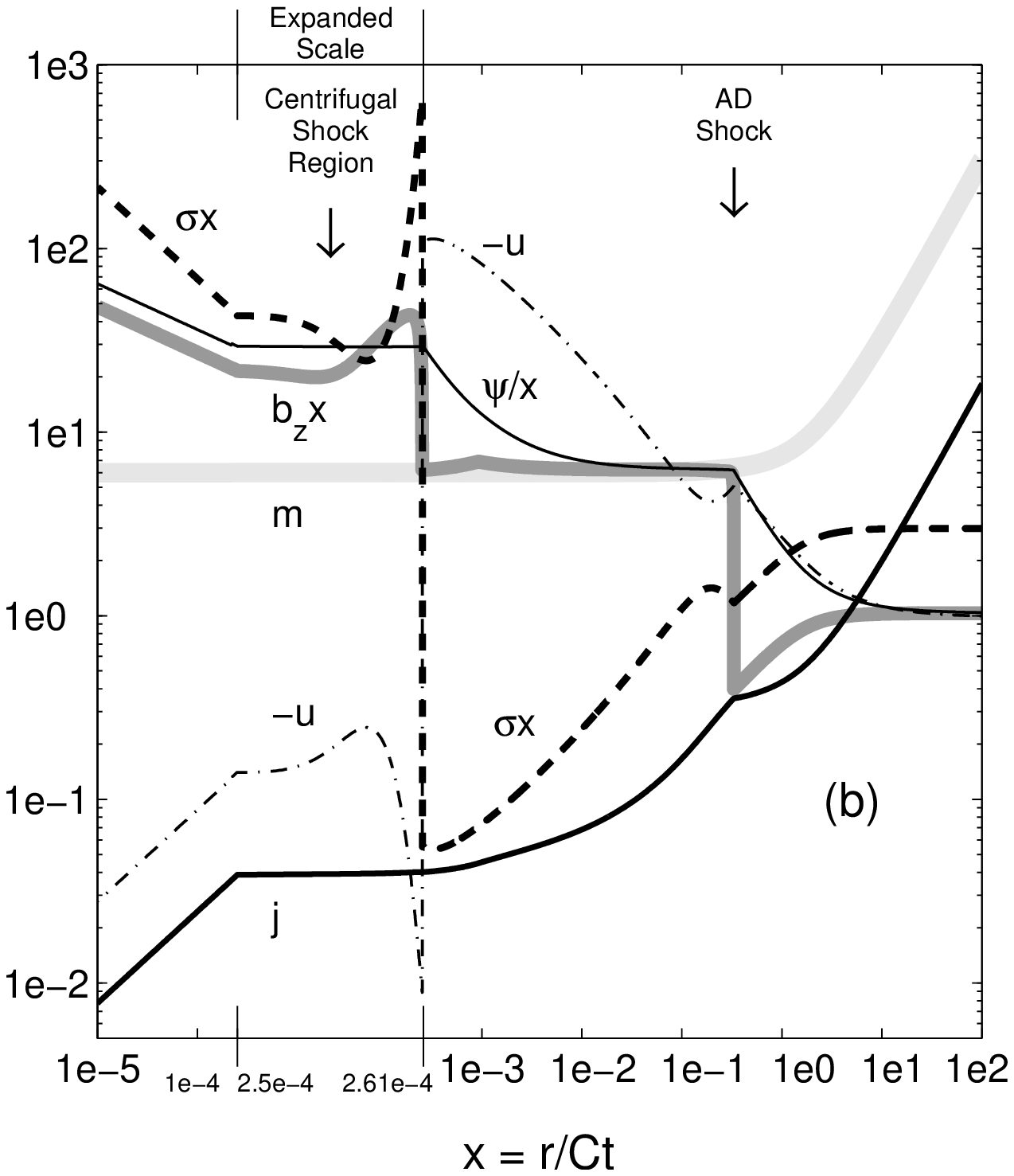}
\figcaption[f8a.eps,f8b.eps]{\label{8}(a) Same parameter
values as in Fig.\ \ref{7}, except that
$v_0$ is reduced from 0.73 to 0.18. In this case $x_a=0.46$,
$x_c=1.7\times{}10^{-3}$, $m_c=6.0$, $\mplateau=6.1$, and
$\jplateau\approx0.1$.
(b) Same as Fig.\ \ref{8}a, except that the diffusivity parameter
$\eta$ is reduced from 1 to 0.7. In this case $x_a=0.33$,
$x_c=2.6\times{}10^{-4}$, $m_c=6.00$, $\mplateau=6.03$, and
$\jplateau\approx0.04$. The horizontal scale in the
vicinity of the centrifugal shock has been expanded to show details of the
postshock transition zone.}}\columnbreak{}

{\noindent{}As we point out in
\S~\ref{discussion}, $m_c$ will be reduced below the value of
$m$ upstream of the centrifugal shock if (as expected) the
rotationally supported disk drives a wind from its surfaces.
The numerical value of the reduction factor depends, however, on yet
another model parameter (see eq.\ [\ref{m_cw}]). Therefore, to simplify the
discussion, we have chosen not to incorporate the effect of
the wind
mass loss into the solutions that we present.}

The centrifugal force starts to become important near the
downstream end of this zone: after it comes to exceed the
gravitational force, it triggers the formation of the
centrifugal shock at the point $x_c\approx\jplateau^2/m_c$. In
contrast to the IMHD solution presented in \S~\ref{MHDstandard}, one
cannot approximate $\jplateau$ by its initial value
in this
case: the braking action induced by the magnetic field
amplification in the AD shock completely invalidates this
approximation. A comparison of Figures \ref{8}a and \ref{8}b
illustrates this point.
When $\eta$ is reduced from 1 (Fig.\ \ref{8}a) to 0.7 (Fig.\ \ref{8}b), the
location of the AD shock changes by a factor of $\sim 0.7$
and the value of the central mass remains essentially the same:
this is as expected from equations (\ref{x_a}) and (\ref{m_pl}), respectively.
However, the decrease in $\eta$ has a very pronounced effect on the
location of the centrifugal shock: as $\eta$ is reduced by a
factor of $\sim 1.4$, $x_c$ decreases by a factor of $\sim 6.7$.
As we now demonstrate, this can be traced to the sensitive
dependence of the magnetic braking term on $\eta$, which leads
to a strong reduction in $j$ downstream from the AD shock for
even a moderate decrease in $\eta$.

We first estimate $b_{\phi,s}$ by substituting $b_z$ from
equation (\ref{bhi}) into equation (\ref{bphiequation}) and
using equation (\ref{mw}),
\begin{equation}
\label{bphisapproximation}
b_{\phi,s}\approx -2\alpha\psi jx^{-3}(1+2\alpha w)^{-1}\ .
\end{equation}
Equation (\ref{angmomequation}) then yields
\begin{eqnarray}
\frac{dj}{dx}&\approx&-(x^2/m)b_z b_{\phi,s}\\
\label{djdx}
&\approx&2\alpha\eta^{-1}\, j (\sigma/h)^{1/2}\, (1+2\alpha w)^{-1}\ .
\end{eqnarray}
Using again the approximations $b_{r,s}\approx b_z$ and
$h\approx 2\sigma/b_{r,s}^2$ that were employed in deriving the
above estimate of $x_a$, equation (\ref{djdx}) becomes
\begin{equation}
\frac{dj}{dx}\approx
\frac{\alpha}{\eta^2}\frac{m}{x}\, j\, (1+2\alpha w)^{-1}\ .
\end{equation}
{Using the approximations $w\approx(2m/x)^{1/2}$
and $m\approx m_c$,
and assuming that $2\alpha w\gg 1$, we get}
\begin{equation}
\frac{dj}{dx}\approx
(m_c/8)^{1/2}\eta^{-2}x^{-1/2}\, j\ .
\end{equation}
This equation can be integrated to give
\begin{equation}
\label{jbrakingapprox}
j=\jconst\exp\left\{(m_c/2)^{1/2}\eta^{-2}x^{1/2}\right\}\ ,
\end{equation}
where $\jconst$ is a constant.
This result is valid between $x_a$ and $x_c$ and can be used to
estimate the latter.  For $x_c\ll x_a$, the value
of $j$ at $x_c$ may be identified with $\jplateau$ and can
be approximated by $\jconst$. Assuming also that no significant
magnetic braking takes place for $x>x_a$ (consistent with our
analysis of the IMHD solution in \S~\ref{MHDstandard}), we can write
$\jplateau\approx(v_0/A)\mplateau\approx(v_0/A)m_c$. Using
equation (\ref{x_a}) to evaluate $x_a$, we finally obtain
\begin{equation}
\label{xcADapproximation}
x_c\approx\frac{v_0^2}{A^2}
m_c\exp\left\{-(2^{3/2}\eta^{-3}m_c/\mu_0)^{1/2}\right\}\ .
\end{equation}
We have verified that this estimate (which depends on the ratio
$x_a/x_c$ being $\gg 1$)
yields the value of $x_c$ to within 30\% for both of
the solutions shown in Figure \ref{8}.
The strong dependence of $x_c$ on $\eta$
that is exhibited by these solutions can be understood from the form of
the expression (\ref{xcADapproximation}): it indicates that the
effect of even a small variation in $\eta$ is magnified on
account of its presence inside an exponential function.
Although the precise form of this expression may be expected to
differ in cases where some of the approximations we utilized do
not apply, the sensitive dependence of $x_c$ on $\eta$
is probably a general feature of the model.

\paragraph{The centrifugal shock ($x\approx{}x_c$).}
The shock represents a discontinuity in the variables $u$,
$\sigma$, and $b_z$ through which the flow changes from the
nearly undiluted free fall to subsonic accretion.
We approximate the shock as a discontinuity in the variables $u$
and $\sigma$, which is governed by the isothermal-shock jump
conditions (Appendix~\ref{isothermalshock}), followed by a very
thin (but numerically resolvable) layer where $b_z$ increases. In this picture,
the ions are decoupled from the neutrals inside the subshock, so their
radial speed in the frame of the shock, and hence (by magnetic flux
conservation, see eq.\ [\ref{flux_conserve}]) $B_z$, remain
unchanged across the subshock even as
$|V_r|$ and $\Sigma$ vary. However, the increase in the drift
speed $|V_{D,r}|$ across the subshock enhances the collisional
drag force immediately behind it and causes a local recoupling
of ions and neutrals; this, in turn, leads to a rapid decline in
$|V_{i,r}|$ and a consequent increase in $B_z$.
Just inward of the shock there is a narrow transition
zone where the flow settles down to its asymptotic Keplerian
structure.
Within this layer the variables $u$, $\sigma$, and $b_z$
adjust rapidly (with some oscillations and overshoots) to their
asymptotic values. This implies, in particular, that, just as in
the case of the AD shock, the
derivative terms $h db_z/dx$ in the constituent equations cannot
be neglected in this zone. The surface density in this region is
significantly larger than $\sigma$ in its immediate vicinity,
so the layer represents a massive ring lying just outside the main
body of the disk (resembling the situation in the fiducial IMHD solution
considered in \S~\ref{MHDstandard}). In the solution shown in
Figure \ref{7}, the width of the ring is $\sim 10\%$ of
$x_c$ and it encompasses $\sim 8\%$ of the total mass of the
disk within $x_c$; the
relative size of the ring in the solutions displayed in Figure \ref{8}
is, however, smaller. The variables $\psi$ and $j$
do not change appreciably across this region.
\paragraph{The Keplerian disk ($x<x_c$).}
Inward of the postshock transition layer, the flow approaches
the form of a magnetized Keplerian accretion disk as
described by equations (\ref{asymptoticm})--(\ref{asymptotich}).
For the solutions shown in Figures \ref{7}, \ref{8}a, and \ref{8}b, the
total mass of the rotationally supported disk (including
the ring at $x_c$) is, respectively, $\sim 5\%$, $\sim 2\%$, and
$\sim 0.4\%$ of $m_c$.

\subsubsection{Fast Rotation}
\label{ADfast}
As in the corresponding IMHD case, a high value of $v_0$ and
comparatively low values of
$\alpha$ and $\delta$ result in the centrifugal shock being located
within the self-gravity--dominated region (where $m\gg
m_c$). The value of $x_c$ is large enough for ideal MHD to still
be applicable (so $\psi\approx xwb_z$). Since the inflow speed
is strongly reduced in the dense disk that forms behind the
centrifugal shock, the ambipolar-diffusion time becomes shorter
than the accretion time and an AD transition front is
established not far from $x_c$. In fact, as $-u$ is
small enough for the flow to remain subsonic ($x-u<1$) in this
region, the IMHD-to-AD transition is gradual
rather than sharp as in the other cases considered in this
subsection, where it
takes the form of an AD shock. Since the centrifugal shock
occurs within the ideal-MHD region, we impose the same jump
conditions at $x_c$ as in the corresponding IMHD case (see
\S~\ref{generalizedisothermalshock}).

A representative solution is shown in Figure \ref{9}. It exhibits the
same general features (a non-Keplerian outer region, a small\columnbreak{}
\mbox{\plotone{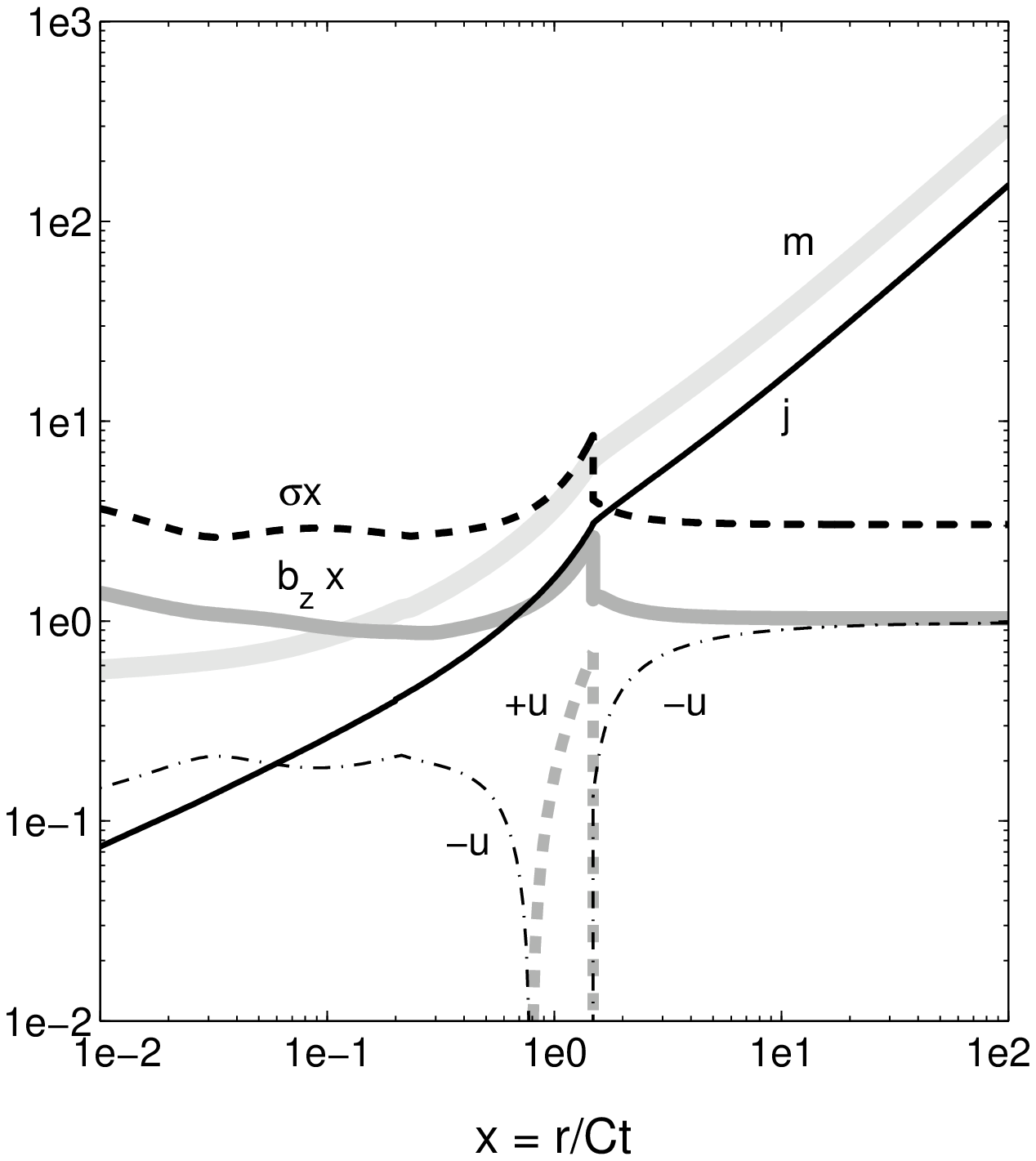}}

{\figcaption[f9.eps]{\label{9}Fast-rotation solution for an
ambipolar-diffusion rotational
collapse. The variations of the normalized radial
infall speed $-u$, surface density
$\sigma$, total mass $m$, specific angular momentum $j$, and $z$
component of the magnetic field $b_z$ are
plotted as functions of the similarity variable $x$. The
model parameter values are the same as in the fiducial case
(Fig.\ \ref{7}), except that the azimuthal-velocity parameter $v_0$ is
increased from 0.73 to 1.5 and the braking parameter
$\alpha$ is increased from 0.08 to 0.1 (so they match the
corresponding parameters in the ideal-MHD solution depicted in Fig.\ \ref{4}).
The derived values of the centrifugal-shock
radius ($x_c=1.5$) and of the central mass ($m_c=0.5$) are
similar to those of the corresponding ideal-MHD solution.}}

\vskip 3ex

{\noindent{}central mass, a backflowing layer behind the centrifugal shock)
as the fast-rotation IMHD solution (see Figs.\ \ref{4}
and \ref{5}). These characteristics are basically a consequence of the
fast initial rotation, and the strong similarity with the
corresponding IMHD case shown in Figure \ref{4} can be understood from
the fact that the centrifugal shock occurs outside the AD regime.}

\subsubsection{Strong Braking}
\label{ADstrong}
In \S~\ref{MHDstrong} we have found that ideal-MHD flows with
strong braking can lose all their angular momentum at a finite
distance from the origin. Here we consider whether a similar situation can
occur in the presence of ambipolar diffusion.

When $j=0$ there are two inner asymptotic solutions that are
mathematically self-consistent and that can be matched to the
outer asymptotic solution by integrating the system of
differential equations.
Correspondingly, two kinds of solutions are found: one that is almost IMHD in
nature, and another that is AD dominated.

The IMHD-like solution resembles the one obtained in
\S~\ref{MHDstrong} in that $b_z$ is everywhere well approximated
by the root $\blo$ of equation (\ref{bzquadraticequation}),
there is no AD shock, and the inner asymptotic solution is given by equations
(\ref{MHDstrongasymptoticbegin})--(\ref{MHDstrongasymptoticend})
(but with $\mu$ possibly differing from $\mu_0$ on
account of the small, but finite, ambipolar diffusion).
Outside the point $x_j$ the azimuthal field component can differ
significantly from its IMHD value, but once complete braking is
accomplished and both $v$ and $b_{\phi,s}$ effectively vanish,
this solution basically describes an
IMHD nonrotating inflow and, in particular, predicts the
formation of\columnbreak{}
\mbox{\plotone{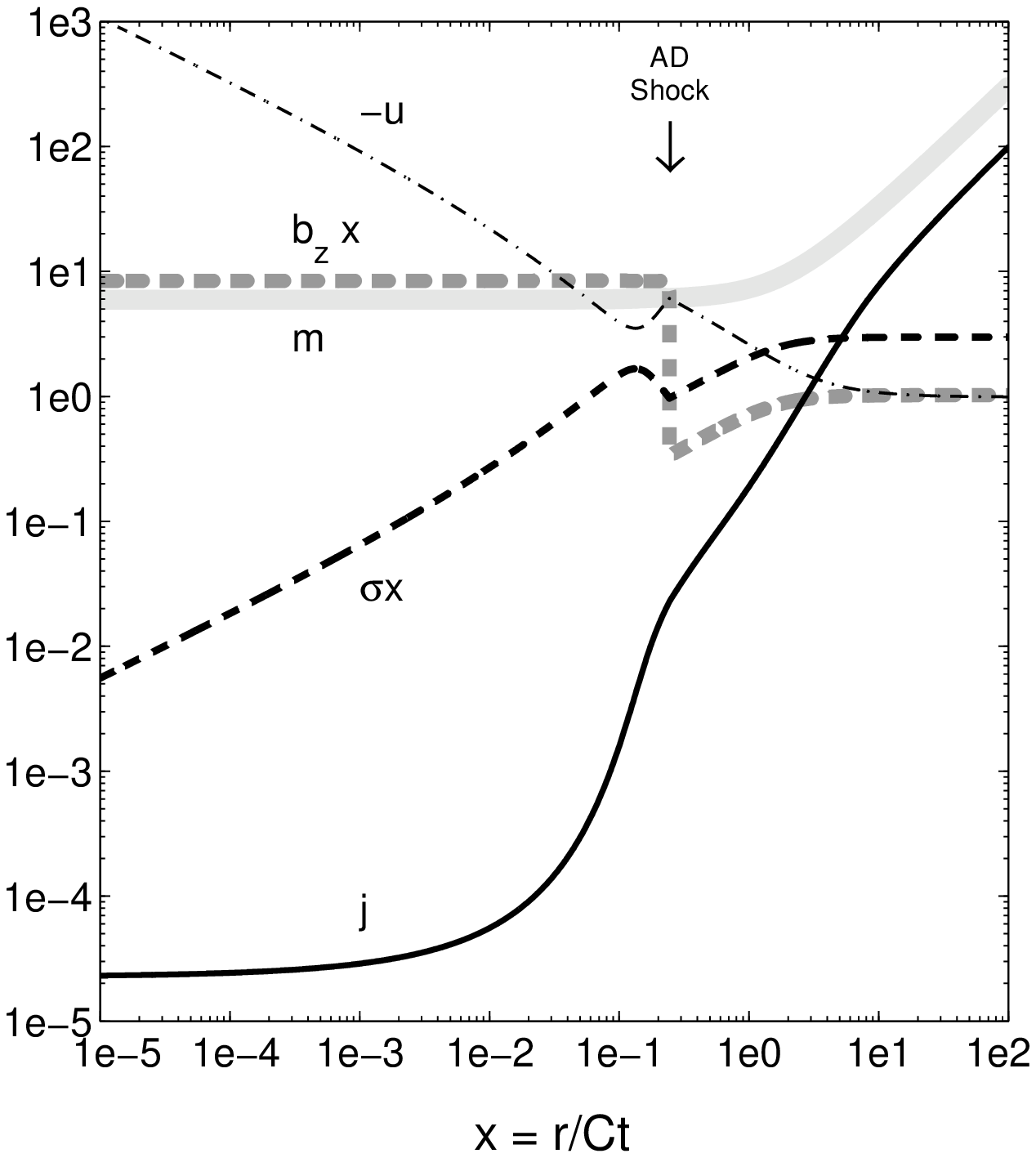}}

{\figcaption[f10.eps]{\label{10}Strong-braking
solution for an ambipolar-diffusion rotational
collapse. The variations of the normalized radial
infall speed $-u$, surface density $\sigma$, total mass $m$,
specific angular momentum $j$, and $z$ component of the magnetic
field are
plotted as functions of the similarity variable $x$.
The model parameters are $\eta=0.5$, $v_0=1$,
$\alpha=10$, and $\delta=10$, with the remainder having the same
values as in the fiducial case (Fig.\ \ref{7}).
Note that $j$ attains a tiny, but finite, plateau value in the
innermost (free-fall) region ($\jplateau = 2 \times10^{-5}$).
The AD shock is located at $x_a=0.24$, and the central
mass is $m_c=5.9$.}}

\vskip 6ex

{\noindent{}a central split-monopole field. As we remarked
at the beginning of \S~\ref{AD},
such a configuration is unlikely to occur in real systems, so we
henceforth concentrate on the more plausible alternative case.}

In the diffusive solution, the flow passes through an AD shock
at a point $x_a$ (which can be estimated by eq.\ [\ref{x_a}])
that
typically exceeds $x_j$. Inward of the ambipolar-diffusion shock,
the poloidal field components have the same form as in the
fiducial solutions, $b_{r,s}\approx b_z\approx\bhi$. As in the
corresponding IMHD case, the inner asymptotic solution satisfies
$j=b_{\phi,s}\approx 0$ and describes
a free fall onto the central mass, which, however, in this case
is {\em not} magnetically diluted but rather resembles the situation in
the AD-dominated infall region of the fiducial solutions. It is given by
\begin{eqnarray}
\label{ADstrongasymptoticbegin}
\dot m = m&=&m_c\ ,\\
-u&=&(2m_c/x)^{1/2}\ ,\\
\sigma&=&(m_c/2x)^{1/2}\ ,\\
\label{ADbzstrong}
b_z&=&2^{-1/8}m_c^{5/8}\eta^{-1/2}{\tilde h}^{-1/4} x^{-1}\ ,\\
\label{ADpsistrong}
b_{r,s}=\psi/x^2&=& b_z\ ,\\
h&=&\tilde h x^{3/2}\ .
\label{ADstrongasymptoticend}
\end{eqnarray}
The power-law scalings of the flow variables with $x$ are the
same as in the asymptotic solution obtained by CCK for the nonrotating AD
collapse. However, because of the magnetic pinching effect
included in the above solution, the expression for the
coefficient $\tilde h$ is more involved:
\begin{eqnarray}
{\tilde h}&=& \left[\frac{(12 D/\sqrt{G} - G)^{1/2} - \sqrt{G}}{2
\sqrt[3]{6}}\right]^2\ ,\\
D&=&N^{1/4}/\eta\ ,\\
N&=&2/m_c\ ,\\
G&=&(2 F)^{1/3} - 8 N (3/F)^{1/3}\ ,\\
F&=&9D^2+[3(27 D^4 + 256 N^3)]^{1/2}\ ,
\end{eqnarray}
with the expression for $\tilde h$ representing the real, positive
solution of the quartic equation
${\tilde h}^2+D{\tilde h}^{1/2} - N=0$.

An example of a strongly braked solution is shown in Figure \ref{10}. A curious
feature of the flow is the value of $j$ as $x\rightarrow 0$,
which assumes a nonzero (albeit rather small) plateau value
($\jplateau\sim 10^{-5}$). This puzzling behavior can be attributed to the very
small magnitude of $\psi$ in this limit, which
causes $|b_{\phi,s}|$ (eq.\ [\ref{bphiequation}]) and therefore
$|dj/dx|$ (eq.\ [\ref{angmomequation}]) to be small, more than
compensating for the increase in $b_z$ behind the AD shock. The
low value of the angular-momentum derivative in turn accounts
for the asymptotic near-flatness of the $j$ curve.
(In contrast, the large central value of $\psi$ associated with the
split-monopole field in the near-IMHD solution guarantees that,
in this case, $|b_{\phi,s}|$ and $|dj/dx|$ remain large enough
near $x_j$ to make $j$ actually vanish there.)
The value of $\jplateau$ is highly sensitive to the magnitude of
the parameter $\eta$: it decreases by $\sim 10$ orders of
magnitude when $\eta$ is decreased from 0.5 to 0.3. (The strong
sensitivity of $x_c$ to the value of this
parameter has already been noted in connection with the fiducial
solutions; see eq.\ [\ref{xcADapproximation}].) However,
for practical purposes, $\jplateau$ can be regarded as being
effectively zero. In particular, the implied
location of the centrifugal radius, $x_c\approx\jplateau^2/m_c$,
corresponds to such a small radius for any meaningful value
of $t$ that it would invariably lie well inside the central protostar.

\section{Discussion}
\label{discussion}
The model presented in this paper provides a physical
framework for a detailed study of the development of star--disk
systems in the collapse of rotating, magnetic molecular-cloud cores.
Perhaps the most noteworthy feature of the solutions presented
in \S~\ref{results} is that, for representative values of the initial
cloud rotation and magnetization and with plausible estimates of the
magnetic field diffusivity and of the strength of magnetic
braking, they predict the formation of circumstellar disks whose
properties compare quite well with those of real protostellar systems.
In particular, the disks obtained for our chosen fiducial
parameters are {\em Keplerian}, having a mass that is only a small
fraction ($\lesssim 10\%$) of the central mass, as has been
inferred for most protostellar disks in nearby dark clouds (see
\S~\ref{formulation}). For the
representative AD solution shown in Figure \ref{7} (corresponding to $\eta=1$,
$v_0=0.73$, $\alpha = 0.08$, and $\delta=1$), the disk mass
$M_d$ is less than $5\%$ of the central mass $M_c$. For
comparison, if the model parameters are slightly changed (to $\eta=1$,
$v_0=0.77$, $\alpha = 0.11$, and $\delta=0.7$),
one finds $M_d/M_c \approx 9\%$.
The most reliable protostellar disk velocity curves have been
obtained in Class II YSOs (i.e., T Tauri
stars, of typical ages $\sim 10^6\, \yr$), which, in contrast
to younger protostars, are no longer strongly obscured by
dusty envelopes (identified with the contracting cloud
cores). These are also the systems with the most accurate
determinations of the disk mass \citep[e.g.,][]{Mundy2000}: the
massive envelopes surrounding Class 0 YSOs prevent their disk
masses from being pinned down, and even in the case of Class
I sources the values of $M_d$ are not as well determined
(although they appear to be similar to the disk masses in Class
II systems). It has occasionally been suggested in the literature that
protostellar disks may not be Keplerian during the early (in
particular, Class-0) phase of their evolution, when the ratio of
the central mass to that of the surrounding envelope is still
small. Our results demonstrate that this need not be the
case. In fact, in the self-similar solutions that we derive, if
the mass of the rotationally supported disk is small in
comparison with the central mass at any given time, it will be so at all times.

The observed rotationally supported disks typically
have sizes $\lesssim 10^2\, \au$ (see \S~\ref{introduction}),
which are reproduced by our typical solutions. For example, the
derived disk size (given by the value of the centrifugal radius $r_c$)
$10^5\, \yr$ after the start of the collapse (which corresponds to
the Class-I evolutionary phase) is $\sim 52\, \au$ for
the fiducial solution presented in Figure \ref{7} and $\sim 130\, \au$
for the other AD solution mentioned in the preceding paragraph.
There have been reports in the literature of T Tauri
circumstellar disks that exhibit Keplerian rotation on somewhat
larger scales [e.g., GM Aur, with $r_c > 200\, \au$ \citep{Dutrey1998},
and DM Tau, with $r_c >600\, \au$
\citep{Guilloteau1998}. It is quite possible that
these disks have achieved their large sizes already during their
early evolutionary stages. For example, they may have possessed
a comparatively high value of the diffusivity parameter $\eta$,
to which the predicted magnitude of $r_c$ is particularly
sensitive (see eq.\ [\ref{xcADapproximation}]). Alternatively,
they may have formed from collapsing cores with comparatively large
initial rotations
and relatively inefficient braking
(see \S~\ref{ADfast}). It is worth noting, though, that in the self-similar
model, disks with larger
values of $x_c$ tend to have higher disk-to-star mass ratios and
therefore exhibit progressively greater departures from a
Keplerian rotation law.
It is, however, also conceivable
that at least part of the cause for large observed disk sizes
in these highly evolved (and no longer strongly accreting)
systems is post-formation viscous spreading of the disk (e.g.,
\citealt{LinPringle1990}; \citealt{Hartmann1998}). It is interesting
to note in this connection that the radial scaling of the disk
surface density predicted by our self-similar solution
($\Sigma\propto r^{-3/2}$ for small $r$; see eq.\ [\ref{ADsigma}]) is the same
as that inferred for the ``minimum mass'' solar nebula
\citep{Weidenschilling1977,Hayashi1985}. Although this
agreement is encouraging, it is not
unique to this model: in particular, the same scaling
is predicted for a self-similar {\em Keplerian} disk with ``$\alpha$
viscosity'' \citep{Tsuribe} and for a self-similar {\em self-gravitating}
disk with a gravitational instability-induced effective viscosity
\citep{LinPringle1987}.

Our model disks exhibit a surface-density enhancement near
their outer edges (see \S~\ref{MHDstandard} and
\S~\ref{ADstandard}): for the fiducial AD solution shown in Figure \ref{7},
this ring has a mass $\sim 0.1\, M_d$ and a width $\sim 0.13\,
r_c$. We are not aware of any observational evidence as yet for
the existence of such rings in protostellar systems, but there have
been several recent reports of well-defined dust rings with similar
radii ($\sim 10^2\, \au$) and fractional widths around older, Vega-like
stars \citep*[e.g.,][]{Schneider1999,Dent2000,Koerner2001}.
Although the apparent confinement of the latter
rings is usually attributed to the presence of solid ``satellites''
(e.g., planets), it is interesting to speculate that their
origin could be traced to the $\Sigma$ enhancements predicted by
our solutions.

An important implication of the result that typical rotating
inflows tend to produce a central mass that is surrounded by a
Keplerian disk is that the angular momentum problem (see
\S~\ref{introduction}) for the modeled YSO is basically
resolved. In particular, angular momentum transport can be
sufficiently efficient to allow most
of the inflowing mass to end up (with effectively no angular momentum) at
the center, with the central mass dominating the dynamics well
beyond the outer edge of the disk even as the inflow is still in
progress. Our solutions reveal that the AD shock,
even though it is usually located well outside the region where
the centrifugal force becomes important, helps to enhance
the efficiency of angular momentum transport through
the magnetic field amplification that it induces. In practice,
the ultimate value of the protostellar angular momentum would be
determined by the braking mechanism that enables the YSO to
reduce its angular velocity below the centrifugal-equilibrium value that
characterizes the accreted disk matter
\citep[e.g.,][]{Mestel1987,K91,Popham1996}.

The formation of Keplerian circumstellar disks in our
``standard'' solutions is a consequence of efficient magnetic
braking, which allows the mass injected at large distances to
reach the center rather than pile up in the disk (as it does in
the fast-rotation solutions discussed in \S~\ref{MHDfast} and
\S~\ref{ADfast}). The disk in these cases therefore has little influence on the
mass accretion rate onto the central YSO, $\dot M_c = 1.59\times
10^{-6}(T/10\, \kelvin)^{3/2} m_c\ M_\odot\, \yr^{-1}$, which is
determined on scales much larger
than $r_c$. In fact, as we discuss in \S~\ref{nonmagnetic},
the eigenvalue $m_c$ is typically also insensitive to the
magnetic structure of the flow and is often well approximated by the
``plateau'' value $\mplateau$ (eq.\ [\ref{m_pl}]) that depends
only on the initial conditions. (For our fiducial parameters,
$\mplateau = 6$, implying $\dot M_c \approx 9.5
\times 10^{-6}\ M_\odot\, \yr^{-1}$.)
The likely production of a centrifugally driven
wind that transports mass away from the disk surfaces (see
discussion below) will, however, act to reduce the value of
$m_c$: as we estimate in Appendix~\ref{wind}, the reduction
factor could be as large as $\sim 3$.

The presence of rotation also does not strongly
affect the ``revitalization'' of ambipolar diffusion behind the AD
shock, which, as we noted in \S~\ref{introduction}, can go a long way toward
resolving the magnetic flux problem. This conclusion
applies even in the fast-rotation case (see \S~\ref{ADfast}),
where the AD shock is replaced by a more gradual
transition occurring within the disk. The asymptotic
$x\rightarrow 0$ similarity solution for the disk implies that, at any
given time, the magnetic flux scales as $r^{3/4}$ (see
eq.\ [\ref{ADpsistandard}]), which represents a slower decrease with
$r$ than in the nonrotating case (in which $\Psi\propto r$; see
eq.\ [\ref{ADpsistrong}]). However, in reality, the amount of magnetic flux
that ends up threading the protostar would depend on the detailed flux
transport and destruction mechanisms that operate in the inner disk and the
star, which may involve Ohmic resistivity
\citep[e.g.,][]{LiMcKee1996}, ambipolar diffusion
\citep[e.g.,][]{DeschMouschovias2001}, and reconnection
\citep[e.g.,][]{GoodsonWinglee1999}.

The quasi-equilibrium disk configurations described by our
solutions make it possible to investigate the stability
properties of the modeled circumstellar disks. Although a full
pursuit of this topic is beyond the scope of this paper, we
comment briefly on the basic MHD and gravitational instabilities
that may affect the disks. The most relevant MHD
instability is evidently the magnetorotational one, which is
expected to develop rapidly in any differentially rotating disk
that is threaded by a subthermal magnetic field (see
\citealt{BalbusHawley1998} for a review). This
instability has been shown to generate turbulence that can
remove angular momentum from the inflowing gas, so the determination
of whether the disk is susceptible to this instability also has
implications to the self-consistency of our identification of
magnetic braking as the dominant angular-momentum transport
mechanism in the disk.

As it turns out, our diffusive Keplerian disk models are by and large
magnetorotationally {\em stable}. Physically, this is because
the coupling between the matter and the field is generally too
weak to allow the instability to grow. Specifically, a necessary
condition for this instability to operate in a weakly ionized, Keplerian
disk is $\tni<\Omega_K^{-1}$, where $\Omega_K$ is the Keplerian
angular velocity \citep{Blaes1994}. Using our
self-similar variables, this criterion can be written as
$(\sigma x^3/m_c h\eta^2)^{1/2}>1$. Since the Keplerian disk region in
all of our typical solutions
nearly coincides with the region where the inner asymptotic solution
(eqs.\ [\ref{asymptoticm}]--[\ref{asymptotich}]) is
applicable, we can use the latter to evaluate the left-hand side
of the above inequality; it is found to equal
$2/3\delta$.
[The asymptotic solution satisfies the weak-field condition
$V_{Az}^2/C^2<1$ postulated in the derivation of the instability
criterion; indeed, the left-hand side of
the latter inequality (which equals $b_z^2h/\sigma$) vanishes
($\propto x^{1/2}$) as $x\rightarrow 0$.] But the parameter
$\delta$ (defined in \S~\ref{formulation}) is expected to be $\gtrsim
1$, and therefore the instability condition $\tni\Omega_K<1$ is
not satisfied.
Interestingly, our fiducial IMHD solution is also
magnetorotationally stable. In this case the strong magnetic squeezing
prevents the unstable perturbations from fitting inside the
disk. Formally, one requires the critical wavelength $\lambda_c
\approx 3.6 (V_{Az}/\Omega_K)$ to be less than the disk
thickness $2H$ for the disk to be unstable.
Using the inner asymptotic solution given by equations
(\ref{MHDasymptoticbegin})--(\ref{MHDasymptoticend}), with the
help of which we first verify that $V_{Az}^2/C^2$ again tends to
zero as $x\rightarrow 0$, this condition translates into
$1.8(m_c/2\mu^4 x)^{1/2}<1$. For our typical parameters this requires
$x\gtrsim 0.14$, but this lower bound exceeds $x_c$ (the approximate
disk outer boundary) for this solution.

The above results support the choice of magnetic braking as the main
angular momentum transport mechanism in our disk model. This mechanism
remains effective even in the presence of ambipolar diffusion because
the neutral--ion collision time $\tni$, while being comparatively
long, is shorter than the magnetic braking time $\tbr$. In a Keplerian
accretion disk $\tbr \approx r/|V_r|$ and $V_r$ is typically much
smaller than the Keplerian speed; therefore $\tni/\tbr$ can be $\ll 1$
even if $\tni\Omega_K$ exceeds 1. This conclusion continues to
hold when the angular momentum transport in the disk (and hence
the accretion time $r/|V_r|$) is
controlled by a centrifugally driven wind rather than by the
propagation of torsional Alfv\'en waves. 
As we show in Appendix~\ref{wind}, the basic formalism of our model is
unchanged in this case, and the parameter
$\delta$ (which is now determined directly from the centrifugal
wind solution) is again typically small enough for the
linear stability criterion $\tni\Omega_K>1$ to be satisfied
over most of the disk column density. The argument against
instability is reinforced by indications from numerical
simulations \citep[e.g.,][]{HS98} that MRI-driven turbulence and
the resulting angular-momentum transport are significant only
when $\tni\Omega_K\ll 1$. We note in this connection that
even a disk in which the bulk of the material is weakly coupled
is expected to possess well coupled ($\tni\Omega_K<1$)
surface layers from which the centrifugal wind is launched
(see Appendix~\ref{wind}). 
It is, however, likely that the
critical wavelength for instability exceeds the
density scale height in the strongly coupled regions (see \citealt{WK} and
\citealt{Li96}), so the surface layers should typically also be
magnetorotationally stable.

The basic gravitational instability that we consider involves
fragmentation induced by the disk self-gravity. The relevant
stability condition against axisymmetric perturbations is given
by the Toomre criterion $\QToomre\equiv C\kappa/\pi G\Sigma>1$ \citep{Toomre},
where $\kappa=[r^{-3}(dJ^2/dr)]^{1/2}$ is the epicyclic
frequency. Although this criterion is modified somewhat by
magnetic effects (\citealt{ShuLiIsopedic}; R. Krasnopolsky \&
A. K\"onigl, in preparation), the corrections remain small for
nearly Keplerian disks, especially in the AD regime, and we
therefore neglect them here. We have verified that the Toomre
criterion is well satisfied for most of our rotationally
supported disk solutions, the only exceptions being the outermost (backflow)
regions of the fast-rotation configurations (Figs.\ \ref{4},
\ref{5}, and \ref{9}),
where the effect of self-gravity is comparatively large.
This result is significant in that it suggests that in typical
cases protostellar disks may not fragment even during the early
phases of their evolution, when the central mass is still
small. In our model this behavior can be attributed
to the fact that, for representative parameters, the central
gravity already dominates at the location of the centrifugal
shock, so that the disks are nearly Keplerian from the
start. Inasmuch as the inequality $\QToomre\lesssim 1$ signals the onset of
self-gravity--induced angular-momentum transport in the disk
\citep[e.g.,][]{LinPringle1987}, this result also serves to justify
the neglect of such a transport mechanism in our model. We note,
however, that the collapsing core could in principle be unstable
to fragmentation before point-mass formation, or, following PMF,
in the region outside the rotationally supported disk
\citep[e.g.,][]{Boss2000}.
It is also conceivable that
fragmentation might be initiated even before a dynamically
unstable core is formed (possibly triggered by an AD-mediated
instability in the magnetically supported parent cloud; see
\citealt{Zweibel1998} and \citealt{Indebetouw2000}).

Another potential angular-momentum transport mechanism that was
not included in our model involves centrifugally driven disk
outflows. For a cold, thin, Keplerian disk that is threaded by
open magnetic field lines, such an outflow could be launched if
the angle between the meridional projection of the magnetic
field and the rotation axis exceeds $30^{\circ}$ \citep{Blandford1982}.
This condition is satisfied by our asymptotic disk
solutions --- in particular, the AD inner asymptotic solution implies
a radially constant inclination angle $\sim 53^{\circ}$
(see eq.\ [\ref{ADpsistandard}])--- so our model disks can be
expected to give rise to outflows of
this type. This result is encouraging in view of the fact that
centrifugally driven disk outflows are a leading candidate for
the origin of the bipolar outflows that are frequently
observed to emanate from YSOs (see, e.g., \citealt{KP2000}
for a review). A centrifugally driven disk outflow can be incorporated into
our model by taking account of the fact that the wind would be
quasi-steady on the accretion time scale and should thus be well
approximated by a steady-state formulation. The condition that the centrifugal
outflow control the vertical angular momentum transport from the
disk is equivalent to the requirement that the steady-state wind
solution pass through the Alfv\'en critical point. This
requirement, in turn, serves to fix the value of $B_{\phi,s}$
(or, equivalently, of the parameter $\delta$) in the
disk solution. Besides angular momentum, the wind also carries away mass,
resulting in a progressive decline in the mass accretion rate with decreasing
disk radius and leading to lower values of $\dot m$ and $m$ at the center.
Interestingly, the radial scaling of the disk magnetic
field ($\propto r^{-5/4}$; see eq.\ [\ref{ADbz}]) implied by the
asymptotic AD disk solution for any given value of $t$ is the same as in the
\citet{Blandford1982} radially self-similar, steady-state wind
solution. This suggests that the latter solution can be used in
conjunction with our disk model to fix the values of $\delta$ and
of the $m_c$ reduction factor in the asymptotic disk
solution. We discuss the implementation of this scheme in
Appendix~\ref{wind}. Inasmuch as the incorporation of a disk wind
does not qualitatively change our basic results, we have not
included its effects in the model described in \S~\ref{model} so as not to
unduly complicate the presentation. It is nevertheless important
to recognize that our solutions indicate that the collapse of
rotating, magnetic cloud cores naturally gives rise to disks with magnetic
field configurations that are conducive to the
launching of centrifugally driven winds. This conclusion is
supported by the MHD core-collapse numerical simulations of
\citet{Tomisaka1998,Tomisaka2000,Tomisaka2002}, in which disk
outflows that transport angular momentum through material and
magnetic stresses are observed to form. (The possibility that a
disk formed in the collapse of a magnetized rotating
cloud core could give rise to a centrifugally driven wind was
recognized also by \citet{CS01}. However, in contrast with our
model, in their picture the angular momentum transport in the
disk is dominated by turbulent viscosity and the outflow is
concentrated near the outer disk radius; furthermore, they infer that the
mass outflow rate depends sensitively on the efficiency of
magnetic braking in the precollapse core.)

We have also presented solutions (\S~\ref{MHDstrong} and
\S~\ref{ADstrong}) in which the
magnetic braking is sufficiently efficient to prevent the
formation of a rotationally supported disk from the collapse of
a rotating cloud core. Although we argued that the parameter
combinations characterizing these solutions are not typical, it is
conceivable that such situations could be realized. A possible
indication that this may sometimes be the case is provided by
the detection of slowly rotating YSOs that show no
evidence (from excess near- and mid-IR emission) for the
presence of a circumstellar disk \citep{Stassun1999,Stassun2001}.
These observations are puzzling if one
considers disk accretion to be a ubiquitous feature of star
formation and interprets slow stellar rotation as due to a
star--disk interaction \citep[e.g.,][]{K91,Popham1996}. The
properties of these YSOs are, however, readily understood if
they are the product of a strongly braked core collapse.

\section{Summary}
\label{summary}
We study the collapse of rotating, magnetic
molecular-cloud cores with the help of a
self-similar model. This model generalizes the work of CCK
on nonrotating cloud cores by incorporating the effects of rotation and
magnetic braking (while also taking into account ambipolar diffusion
and its dependence on the magnetic tension force as in the
original model). We focus on the evolution after point-mass
formation, noting that the pre-PMF collapse has already been
studied by numerical
simulations. Our approach is motivated by previous findings that
simulation results for the pre-PMF collapse of rotating cores and
for the pre- and post-PMF collapse of nonrotating cores are well
approximated by self-similar solutions. Our semianalytic scheme
allows us to examine the full range of expected behaviors and
their dependence on the physical parameters. We present and
analyze solutions for rotating but nonmagnetic collapsing cores,
reproducing the results of \citeauthor{SH} (\citeyear{SH}), as
well as for magnetic cores in the ideal-MHD and
ambipolar-diffusion limits. In these two regimes, we distinguish
between fiducial solutions, obtained for typical parameter
values that correspond to moderately fast initial cloud rotation rate
and moderately strong magnetic braking, and limiting cases of (1) fast
rotation and (2) strong braking. Our results can be summarized
as follows:
\begin{itemize}

\item For representative parameter values, we obtain solutions
that describe rotationally supported circumstellar disks of masses and sizes
that are consistent with observations of YSOs. Our model thus
makes it possible, for the first time, to study the formation of
protostellar disks in the context of a realistic scenario of
star formation in magnetically supported, weakly ionized
molecular cloud cores. The outer boundary of the nearly
stationary accretion disk roughly coincides with the location of
the centrifugal shock, which typically occurs inward of the
ambipolar-diffusion shock where the magnetic field decouples from the matter.

\item Our solutions indicate that it is quite possible that
T Tauri (Class II) protostellar systems, whose disk masses are
typically inferred to be $\lesssim 10\%$ of the central mass, have had a
similarly low disk-to-star mass ratio also during their earlier
(Class-0 and Class-I) evolutionary phases.

\item The disk configurations that we obtain exhibit
surface-density enhancements near their outer boundaries
(encompassing $\lesssim 10\%$ of the disk mass and radius for
typical parameters). We speculate that these features may
be the precursors of the rings detected in some Vega-like systems on
scales $\lesssim 10^2\, \au$.

\item Our models allow us to elucidate the interplay between the
ambipolar diffusion and magnetic braking processes. In
particular, the magnetic field enhancement behind the AD shock
increases the efficiency of angular-momentum removal from the
disk. In the limiting case of strong braking, essentially all
the angular momentum is removed well before the inflowing gas
reaches the center. Such systems may correspond to
slowly rotating YSOs that show no evidence of a circumstellar
disk.

\item The inferred magnetic-field structures imply that the
disks could drive centrifugal outflows over much of their radial
extent. We show that the steady-state, radially self-similar wind
solution of \citet{Blandford1982} can be naturally incorporated
into the asymptotic AD disk solution, making it possible to
study the effects of wind angular-momentum and mass removal from
the disk and to better constrain the relevant parameters.

\item The derived nearly-Keplerian disk configurations appear to
be immune to both the magnetorotational instability and to
self-gravity--induced fragmentation. Only in the limiting case
of rapid initial rotation (which yields comparatively
massive and largely non-Keplerian disk solutions) is the Toomre $Q$ parameter
$< 1$ anywhere in the disk.

\end{itemize}

Despite the various simplifications that needed to be employed
to ensure self-similarity, our solutions likely capture the main
qualitative aspects of the collapse of rotating cloud cores, and
it is conceivable that our numerical estimates are also not too far off.
Further progress in the study of protostellar disk formation
could be achieved by means of numerical MHD simulations.
\acknowledgments
We thank J. Everett, W. Herbst, Z.-Y. Li, and N. Vlahakis for helpful
conversations or correspondence. We also acknowledge I. Contopoulos
and G. Ciolek,
whose insights on the nonrotating core-collapse problem (presented in
CCK and CK) have guided us in the present work. This research was
supported in part by NASA grant NAG 5-3687 and by
DOE under grant B341495.

\appendix
\section{Disk Equations}
\label{diskequations}
Assuming axisymmetry and isothermality, we write down (using
cylindrical coordinates) the mass, radial momentum, and angular momentum
conservation relations as well as the vertical hydrostatic
equilibrium condition:
\begin{equation}
\label{mass3d}
\frac{\partial\rho}{\partial t}
+
\frac{1}{r}\frac{\partial}{\partial r}\left(r\rho V_r\right)
=
-\frac{\partial}{\partial z}\left(\rho V_z\right)\ ,
\end{equation}
\begin{equation}
\label{radial3d}
\rho\frac{\partial V_r}{\partial t}
+\rho V_r \frac{\partial V_r}{\partial r}
=
\rho g_r
-
C^2\frac{\partial \rho}{\partial r}
+\rho\frac{V_\phi^2}{r}
+\frac{B_z}{4\pi}\frac{\partial B_r}{\partial z}
-\frac{\partial}{\partial r}\left(\frac{B_z^2}{8\pi}\right)
-\frac{1}{8\pi r^2}\frac{\partial}{\partial
r}\left(rB_\phi\right)^2 -\rho V_z\frac{\partial V_r}{\partial
z}\ ,
\end{equation}
\begin{equation}
\label{angular3d}
\frac{\rho}{r}\frac{\partial}{\partial t}\left(r V_\phi\right)
+
\frac{\rho V_r}{r}\frac{\partial}{\partial r}\left(r V_\phi\right)
=
\frac{B_z}{4\pi}\frac{\partial B_\phi}{\partial z}
+
\frac{B_r}{4\pi r}\frac{\partial}{\partial
r}\left(rB_\phi\right) -\rho V_z\frac{\partial}{\partial
z}(rV_\phi)\ ,
\end{equation}
\begin{equation}
\label{vertical3d}
C^2\frac{\partial\rho}{\partial z}
=
\rho g_z
-
\frac{\partial}{\partial z}
\left(\frac{B_\phi^2}{8\pi}+\frac{B_r^2}{8\pi}\right)
+
\frac{B_r}{4\pi}\frac{\partial B_z}{\partial r}\ ,
\end{equation}
where $g_r$ and $g_z$ are, respectively, the radial and
vertical components of the gravitational field.

We now integrate these equations over $z$.  We assume that the
disk is geometrically thin [half-thickness $H(r) \ll r$] and
aim to get expressions that are valid to order $(H/r)^2$. On
account of the disk thinness, we approximate the density, radial
velocity, azimuthal velocity, and radial gravity
as being constant with height. We thus equate the column density
$\Sigma \equiv \int_{-\infty}^\infty\rho\,dz$ to $2H\rho$ and
drop the last term on the right-hand side in both equations
(\ref{radial3d}) and (\ref{angular3d}). Equation (\ref{mass3d}) becomes
\begin{equation}
\label{mass4d}
\frac{\partial\Sigma}{\partial t}
+\frac{1}{r}
\frac{\partial}{\partial r}
\left(r\Sigma V_r\right)
=
-\frac{1}{2\pi r}\frac{\partial{\dot{M}}_w}{\partial r}\, ,
\end{equation}
where the last term on the right-hand side represents the mass
flux in a disk outflow [of total outflow rate $\dot{M}_w(r)$
within a radius $r$]. We neglect this term in the calculations
presented in the body of the paper, but we return in
\S~\ref{wind} to consider the effect of having $\dot{M}_w \ne 0$.

Using the solenoidal condition on the magnetic field,
[$\lfrac{\partial B_z}{\partial
z}=-r^{-1}({\partial}/{\partial r})(rB_r)$, which in a thin
disk allows us to treat $B_z$ as being constant with height
whenever it is not explicitly differentiated with respect to
$z$] as well as the assumed field symmetry, equation
(\ref{radial3d}) integrates to
\begin{eqnarray}
\label{radial3d1}
\Sigma
\frac{\partial V_r}{\partial t}
+
\Sigma V_r
\frac{\partial V_r}{\partial r}
&=&
\Sigma g_r
-C^2
\frac{\partial\Sigma}{\partial r}
+\Sigma\frac{V_\phi^2}{r}
+\frac{B_z B_{r,s}}{2\pi}
\nonumber\\
& &
\mbox{}
-2H
\frac{\partial}{\partial r}
\left(
\frac{B_z^2}{8\pi}
\right)
+\frac{1}{8\pi r^2}
\int_{-\infty}^\infty
\frac{\partial}{\partial r}
\left[r^2\left(B_r^2-B_\phi^2\right)\right]\,dz \ .
\end{eqnarray}
Taking account of the adopted vertical mass distribution, we replace the
integral in the last term of equation (\ref{radial3d1}) by an an integral
over a finite interval,
$\int_{-H(r)}^{H(r)} dz\,$:
\begin{eqnarray}
\frac{1}{8\pi r^2}
\int_{-H(r)}^{H(r)}
\frac{\partial}{\partial r}
\left[r^2\left(B_r^2-B_\phi^2\right)\right]\,dz
&=&
\frac{1}{8\pi r^2}
\frac{\partial}{\partial r}
\left[r^2
\int_{-H(r)}^{H(r)}
\left(B_r^2-B_\phi^2\right)\,dz
\right]\nonumber\\
& &\mbox{}
-
\frac{1}{4\pi}
\left(B_{r,s}^2-B_{\phi,s}^2\right)\left(\frac{dH}{dr}\right)
\end{eqnarray}
\citep*[see][]{Lovelace1994}.
We assume $B_r(r,z)=B_{r,s}(r)\,\left[z/H(r)\right]$, and similarly for
$B_\phi$. This choice is motivated by the
field configuration derived in the case of a rotationally supported thin disk
in which the field is comparatively well coupled to the matter
\citep[e.g.,][]{WK}. Although the disks obtained in our AD collapse solutions
evidently correspond to the weak-coupling case (see
\S~\ref{discussion} and Appendix~\ref{wind}), for which this
approximation is no longer adequate
\citep[e.g.,][]{Li96,Wardle97}, we nevertheless adopt the above
scalings for the sake of definiteness. We emphasize, however,
that none of the dominant terms in the equations that we solve
depends on the details of the vertical variation of these field components.
We thus get
\begin{eqnarray}
\frac{1}{8\pi r^2}
\int_{-H(r)}^{H(r)}
\frac{\partial}{\partial r}
\left[r^2\left(B_r^2-B_\phi^2\right)\right]\,dz
&=&
\frac{1}{12\pi r^2}
\frac{\partial}{\partial r}
\left[r^2 H(r)
\left(B_{r,s}^2-B_{\phi,s}^2\right)
\right]
-
\frac{1}{4\pi}
\left(B_{r,s}^2-B_{\phi,s}^2\right)\left(\frac{dH}{dr}\right)\nonumber\\
&=&
\frac{H}{12\pi r^2}\frac{\partial}{\partial r}
\left[r^2(r^2 B_{r,s}^2-B_{\phi,s}^2)\right]
-
\frac{1}{6\pi}\left ( \frac{dH}{dr} \right )
\left(B_{r,s}^2-B_{\phi,s}^2\right)\, .
\end{eqnarray}
Approximating $g_r$ by $GM(r)/r^2$ [with $M(r)\approx M_c$ if the central
mass dominates] and substituting $J=rV_\phi$,
the integrated radial momentum equation becomes
\begin{eqnarray}
\label{radial3d2}
\frac{\partial V_r}{\partial t}
+
V_r
\frac{\partial V_r}{\partial r}
&=&
-\frac{GM}{r^2}-\frac{C^2}{\Sigma}
\frac{\partial\Sigma}{\partial r}
+\frac{J^2}{r^3}
+\frac{B_zB_{r,s}}{2\pi\Sigma}
-\frac{HB_z}{2\pi\Sigma}\frac{\partial B_z}{\partial r}
\nonumber\\
& &\mbox{}
+\frac{H}{12\pi\Sigma r^2}
\frac{\partial}{\partial r}
\left(r^2 B_{r,s}^2-r^2 B_{\phi,s}^2\right)
-
\frac{1}{6\pi\Sigma}\left ( \frac{dH}{dr} \right )
\left(B_{r,s}^2-B_{\phi,s}^2\right)\ .
\end{eqnarray}

When integrating the angular momentum equation (\ref{angular3d}),
we write
\begin{equation}
\frac{B_z}{4\pi}\frac{\partial B_\phi}{\partial z}
+
\frac{B_r}{4\pi r}
\frac{\partial}{\partial r}
\left(r B_\phi\right)
=
\frac{1}{4\pi}
\frac{\partial}{\partial z}
\left(B_z B_\phi\right)
+
\frac{B_r}{4\pi r}
\frac{\partial}{\partial r}
\left(r B_\phi\right)
+
\frac{B_\phi}{4\pi r}
\frac{\partial}{\partial r}
\left(r B_r\right)\ ,
\end{equation}
and use the same ansatz as above for $B_r(r,z)$ and $B_\phi(r,z)$.
We then obtain
\begin{equation}
\label{angular3d1}
\frac{\partial J}{\partial t}
+
V_r\frac{\partial J}{\partial r}
=
\frac{rB_zB_{\phi,s}}{2\pi\Sigma}
+
\frac{H}{6\pi r \Sigma}
\frac{\partial}{\partial r}
\left(r^2B_{r,s}B_{\phi,s}\right)
-\frac{r B_{r,s}B_{\phi,s}}{3\pi\Sigma}\left(\frac{dH}{dr}\right)\ .
\end{equation}

In integrating the vertical hydrostatic-balance equation, we
take the pressure at the disk surfaces to vanish. The midplane
pressure $p_c \approx ({\Sigma}/{2H})C^2$ is then given by
\begin{equation}
\label{vertical3d1}
p_c = \frac{\pi}{2}G\Sigma^2
+\frac{GM_c\Sigma H}{4r^3}
+\frac{B_{r,s}^2+B_{\phi,s}^2}{8\pi}
-\frac{HB_{r,s}}{8\pi}\frac{\partial B_z}{\partial r}\ ,
\end{equation}
where the first two terms on the right-hand side constitute our
approximation to $g_z$ and represent,
respectively, the local disk self-gravity and the tidal
squeezing by the central point mass.

The ion equation of motion is approximated by
\begin{equation}
\label{drift3d}
\frac{\rho}{\tni}\Ve_D=
\frac{1}{4\pi}
(\nabla\times\B)\times\B\ ,
\end{equation}
where $\Ve_D\equiv\Ve_i-\Ve_n$; we assume that the disk is
sufficiently weakly ionized that the neutral velocity $\Ve_n$ is
practically indistinguishable from the bulk velocity $\Ve$.  The
radial component of this equation is
\begin{equation}
\frac{4\pi\rho}{\tni}
V_{D,r}=
B_z\left(\frac{\partial B_r}{\partial z}
-\frac{\partial B_z}{\partial r}\right)
-\frac{B_\phi}{r}\frac{\partial}{\partial r}\left(rB_\phi\right)\ ,
\end{equation}
which yields, upon integration over $z$,
\begin{equation}
\label{V_Dr}
V_{D,r}=
\frac{\tni}{2\pi\Sigma}
B_z\left(B_{r,s}-H\frac{\partial B_z}{\partial r}\right)
+
\frac{\tni}{6\pi\Sigma}
\left[
\frac{H}{r}
\left(
B_{r,s}
\frac{\partial B_{r,s}}{\partial r}
-B_{\phi,s}
\frac{\partial B_{\phi,s}}{\partial r}
\right)
+\left(B_{\phi,s}^2-B_{r,s}^2\right)
\left(\frac{dH}{dr}\right)\right]
\end{equation}
(see eq.\ [\ref{radial3d2}]).
Similarly, the azimuthal component of equation (\ref{drift3d}),
\begin{equation}
\label{V_Dphi}
\frac{4\pi\rho}{\tni}
V_{D,\phi}=
B_z\frac{\partial B_\phi}{\partial z}
+\frac{B_r}{r}\frac{\partial}{\partial r}\left( rB_\phi \right)\ ,
\end{equation}
gives, when integrated over $z$,
\begin{equation}
\label{viphi}
V_{i,\phi}=V_\phi +
\frac{{\tni}}{2\pi\Sigma}
B_zB_{\phi,s}
+
\frac{{\tni}H}{6\pi\Sigma r}\frac{\partial}{\partial r}
\left(r^2 B_{r,s}B_{\phi,s}\right)
-
\frac{{\tni}}{3\pi\Sigma}
B_{r,s}B_{\phi,s}r
\left(\frac{dH}{dr}\right)
\end{equation}
(see eq.\ [\ref{angular3d1}]).

The flux conservation equation is obtained from Faraday's law
($\lfrac{\partial\B}{\partial t}=-c\nabla\times\E$) and Ohm's law
in the ambipolar diffusion-dominated regime
($c\E=-\Ve_i\times\B$) together with equation (\ref{drift3d}).
The result is
\begin{equation}\label{flux_conserve}
\frac{\partial\B}{\partial t}
=
\nabla\times(\Ve\times\B)
+\nabla\times
\left\{
\frac{\tni}
{4\pi\rho}
\left[
(\nabla\times\B)\times\B
\right]
\times\B
\right\}\ .
\end{equation}
The $z$ component of this equation is
\begin{eqnarray}
\frac{\partial B_z}{\partial t}
&=&
\frac{1}{2\pi r}
\frac{\partial}{\partial r}
\left(
\frac{\partial\Psi}{\partial t}
\right)\nonumber\\
&=&
\frac{1}{r}
\frac{\partial}{\partial r}
\left(
-rV_rB_z
+
\frac{\tni r}{4\pi\rho}
\left\{
-B_z
\left[B_z
\left(
\frac{\partial B_r}{\partial z}
-
\frac{\partial B_z}{\partial r}
\right)
-\frac{B_\phi}{r}
\frac{\partial}{\partial r}
\left(r B_\phi\right)
\right]
\right.\right.
\nonumber\\
& &\left.\left.\mbox{}
-B_r
\left[
B_\phi
\frac{\partial B_\phi}{\partial z}
+
B_r
\left(
\frac{\partial B_r}{\partial z}
-
\frac{\partial B_z}{\partial r}
\right)
\right]
\right\}
\right)\ ,
\end{eqnarray}
which implies
\begin{equation}
\frac{2}{r}
\frac{\rho}{\tni}
\left(
\frac{\partial\Psi}{\partial t}
+
2\pi rV_rB_z
\right)
=
-(B_z^2+B_r^2)
\left(
\frac{\partial B_r}{\partial z}
-
\frac{\partial B_z}{\partial r}
\right)
+\frac{B_\phi B_z}{r}
\frac{\partial}{\partial r}\left(rB_\phi\right)
-B_rB_\phi
\frac{\partial B_\phi}{\partial z}\ .
\end{equation}
Integrating this over $z$ and rearranging, gives
\begin{eqnarray}
\label{psifull}
\frac{\partial\Psi}{\partial t}
&=&
-2\pi rV_rB_z
+
\frac{r\tni}{\Sigma}
\left\{
-\left(B_{r,s}-H\frac{\partial B_z}{\partial r}\right)
\left(B_z^2+\frac{1}{3}B_{r,s}^2\right)
\right.
\nonumber\\
&+&
\frac{1}{3}B_{\phi,s}^2
\left[
-B_{r,s}
+HB_z\left(
\frac{d}{dr}\ln(rB_{\phi,s})
-
\frac{d\ln H}{dr}\right)\right]
\nonumber\\
&-&
\left.
\frac{1}{3}B_{r,s}^2
\left[
HB_z\left(
\frac{d}{dr}\ln(rB_{r,s})
-
\frac{d\ln H}{dr}\right)\right]\right\} \ .
\end{eqnarray}

In formulating the self-similar model in \S~\ref{basiceqns} we
use a pared-down version of the above equations. This is
motivated by our finding that our basic results can be derived {\em without}
including any $\mathcal{O}(H/r)$ terms. The only
instance in which a term of this order plays a role in our
solutions is in allowing us to refine the structure of the
ambipolar-diffusion shock by employing the
radial gradient of $B_z$ in the radial momentum equation. We
therefore retain the combination $[B_{r,s}-H({\partial
B_z}/{\partial r})]$ (which is proportional to the vertically
integrated azimuthal component of the current density) in
equation (\ref{radial3d2}) but neglect the other two
$\mathcal{O}(H/r)$ terms in that equation, since they typically do not exceed
the $\propto \partial B_z/\partial r$ term.
[The last two terms in equation (\ref{radial3d2}) can become much larger than
the $\propto \partial B_z/\partial r$ term in the limit $r/t \rightarrow 0$
of the IMHD solution, when $B_{r,s}/B_z$ formally
diverges. However, as we argue in \S~\ref{results}, this divergence
is not expected to occur in a real disk. Furthermore, the neglect of
these terms is justified even in our formal solution since $H \propto
1/B_{r,s}^2$ in this limit, so the radial derivative terms remain
negligible in comparison with the magnetic tension ($\propto
B_zB_{r,s}$) term.]
Since the same integrated current density
component appears also in equations (\ref{V_Dr}) and
(\ref{psifull}) for $V_{D,r}$ and $\partial \Psi/\partial t$, respectively,
we keep the combination $[B_{r,s}-H({\partial B_z}/{\partial
r})]$ in these equations too for self-consistency,
but we again omit all the other $\mathcal{O}(H/r)$ terms.
For the same reason we keep the $\propto H(\partial B_z/\partial r)$
term also in the vertical force balance equation
(\ref{vertical3d1}), which means that we end up retaining all the
$\mathcal{O}(H/r)$ terms in this ``master equation'' for $H$.

We similarly neglect the $\mathcal{O}(H/r)$ terms
in the angular momentum equation (\ref{angular3d1}) and,
correspondingly, in equation (\ref{viphi}) for $V_{i,\phi}$. The latter
equation can then be substituted into equation (\ref{BM})
to yield a purely algebraic expression for $B_{\phi,s}$.

\section{Singular Lines and Shock Jump Conditions}
\label{singular}
The singular lines of the flows considered in this paper can be
obtained by rewriting the radial momentum equation
(\ref{forceequation}) in the form
\begin{equation}
\label{shockforceequation}
(1-w^2)\frac{d\sigma}{dx}=-b_z h \frac{db_z}{dx}+[...]\ .
\end{equation}
In the AD limit, in which the field is effectively decoupled from
the bulk of the matter, we immediately infer that the
singular line corresponds to the sonic speed:
\begin{equation}\label{singularAD}
(x-u)^2 = 1\ \ \ \ \ {\text{(AD limit)}}
\end{equation}
(see CCK for a discussion of the physical basis of this result).

In the ideal-MHD limit, flux freezing implies
$b_z/\sigma=1/\mu_0$ (see eq.\ [\ref{initialconditions}]),
and therefore the
singular line is given by
\begin{equation}
\label{singularIMHD}
(x-u)^2=1+h\sigma/\mu_0^2\ \ \ \ {\text{(IMHD limit)}}\ .
\end{equation}
In the region where self-gravity dominates, $h\approx 2/\sigma$, and
equation (\ref{singularIMHD}) reduces to
$(x-u)^2=1+2/\mu_0^2$, which is the result given by equation
(41) in CCK.\footnote{Equation (41) in CCK is therefore only
a special case of the more general expression, which is given by
equation (\ref{singularIMHD}).}
In dimensional form,
equation (\ref{singularIMHD}) is
\begin{equation}
(r/t-V_r)^2=C^2+B_z^2 H/2\pi\Sigma=C^2+B_z^2/4\pi\rho\ ,
\end{equation}
where the last term on the right-hand side is equal to the Alfv\'en speed in
the disk. The IMHD singular line thus corresponds to the
fast-magnetosonic speed, which agrees with (and generalizes) the
statement made in CCK on the basis of the
self-gravity--dominated limit of equation (\ref{singularIMHD}).

When the curve describing the flow crosses and falls below the
singular line in the $\{x,u\}$ plane, this generally signals
the appearance of an unresolved shock discontinuity. In the
present work, this is the situation invariably encountered in the
vicinity of the centrifugal radius. To properly model this
discontinuity, appropriate shock jump conditions need to be
applied. The conservation of mass across the
shock yields
\begin{equation}\label{massflux}
\sigma w_s = {\text{const}}\, ,
\end{equation}
where $w_s \equiv x_s - u$ (with $x_s$ denoting the shock location).
When the ions and neutrals are well coupled inside the shock,
the magnetic-flux conservation condition takes the form
\begin{equation}\label{magneticflux}
b_z w_s  = {\text{const}}\, .
\end{equation}
In the strong-coupling case, equations (\ref{massflux}) and
(\ref{magneticflux}) imply $\mu \equiv \sigma/b_z
={\text{const}}$ across the shock: in what follows we set
$\mu=\mu_0$, which applies when the entire flow
upstream of the shock obeys ideal MHD.
Under the adopted isothermal approximation, the jump conditions
are fully specified once we integrate also the differential
terms in equation (\ref{shockforceequation}) across the
discontinuity. Performing this integral requires making
some assumptions about the behavior of $\sigma$, $b_z$, and $h$
in the shock region. In the following subsections, we consider
four cases in which algebraic jump conditions can be
derived. These results have enabled us to explicitly model the centrifugal
shock in all the core-collapse solutions presented in this paper.
\subsection{Isothermal Shock}
\label{isothermalshock}
If one can neglect variations in the magnetic field across the discontinuity,
as is the case when the centrifugal shock occurs deep inside the AD
regime (see \S~\ref{ADstandard}), equations
(\ref{shockforceequation}) and (\ref{massflux}) imply
\begin{equation}\label{isojump}
\sigma (w_s^2 + 1) = \text{const}\, .
\end{equation}
Denoting the upstream and downstream sides of the shock by the
subscripts 1 and 2, respectively, one can use equations
(\ref{isojump}) and (\ref{massflux}) to express the nontrivial
($w_{s1}\neq w_{s2}$) solution in the form $w_{s1} w_{s2} = 1$. This is the
simplest realization of the ``isothermal'' jump conditions
discussed by \citet{ShuLiIsopedic}.

\subsection{Generalized Isothermal Shock}
\label{generalizedisothermalshock}
If the shock occurs in the self-gravitating regime, where
$h\approx 2/\sigma$, then $b_z h \propto b_z/\sigma$ is constant across the
shock, and the integration of equation (\ref{shockforceequation}) leads to
\begin{equation}
\sigma (w_s^2 + \Theta) = \text{const}\, ,
\end{equation}
where $\Theta\equiv 1+2/\mu_0^2$. This situation is
encountered in the fast-rotation solutions presented in
\S~\ref{MHDfast} and \S~\ref{ADfast}.
In this case the upstream
and downstream radial velocities (in the shock frame) are related by
$w_{s1} w_{s2} =  \Theta$, which constitutes another realization of
the ``isothermal'' jump conditions.
remark.
[Note that, as
$w_{s2}\rightarrow w_{s1}$, one recovers the singular-line relation
(eq.\ [\ref{singularIMHD}]) for a self-gravitating IMHD flow.]

\subsection{Magnetically Squeezed Shock}
\label{magneticshock}
If magnetic squeezing due to the radial field dominates gravity in the
vertical force-balance equation, then $h\approx 2\sigma
b_{r,s}^{-2}$ (see eq.\ [\ref{MHDasymptoticend}]), and equation
(\ref{shockforceequation}) yields
\begin{equation}
\sigma(w_s^2+1) + \frac{2\sigma^3}{3\mu_0^2b_{r,s}^2} = \text{const}\, .
\end{equation}
Taking into account the relation (\ref{massflux}), we obtain a quartic
equation for $\sigma_2$ (whose coefficients are combinations of the
upstream flow
variables). We can eliminate the trivial solution
$\sigma_2=\sigma_1$ by dividing this equation by
$(\sigma_2-\sigma_1)$, which reduces it to
\begin{equation}
\label{shockcubic}
\sigma_2^3+\sigma_1\sigma_2^2+\left(\sigma_1^2+Q\right)\sigma_2-
\sigma_1(u_1-x_s)^2Q=0\ ,
\end{equation}
where $Q\equiv3\mu_0^2b_{r,s}^2/2$.
Equation (\ref{shockcubic}) is a cubic of the form
$\sigma_2^3+a_2\sigma_2^2+a_1\sigma_2+a_0=0$, whose discriminant is
$D=\left(p/3\right)^3+\left(q/2\right)^2$, with
$p=a_1-a_2^2/3$
and $q=a_0-a_1a_2/3+2a_2^3/27$.
For this particular cubic, the discriminant is always positive,
guaranteeing that there is exactly one real root, given by
\begin{equation}
\sigma_2=-a_2/3 + q_{+}+q_{-}\ ,
\end{equation}
where $q_\pm \equiv (-q/2 \pm \sqrt{D})^{1/3}$. This solution
provides a good approximation to the centrifugal shock in the
fiducial IMHD flow (\S~\ref{MHDstandard}).

\subsection{Constant-Thickness Shock}
Another possibility is that $h$ is constant across the shock, as
would be the case if the central point-mass gravity controlled
the squeezing in the vertical equilibrium equation (see
eqs.\ [\ref{asymptotich}] and [\ref{ADstrongasymptoticend}]).
In this case, integration of equation (\ref{shockforceequation}) results in
\begin{equation}
\sigma(w^2+1) + b_z^2 h/2 = \text{const}\, .
\end{equation}
In conjunction with the mass conservation equation
(\ref{massflux}), this can be written as a
cubic equation for $\sigma_2$. After factoring out the trivial
solution, one obtains a quadratic equation,
\begin{equation}\label{quadratic}
\sigma_2^2+(\sigma_1+2\mu_0/h)\sigma_2-(m/x_s)^2(2\mu_0/h\sigma_1)
=0\ ,
\end{equation}
which has exactly one positive real root. In order for this
result to be applicable to a centrifugal shock in a collapsing
magnetized core, the shock must occur far enough from the center
that ideal MHD is still a good approximation (so that the
magnetic term in eq.\ [\ref{shockforceequation}] continues to play a
role). However, in the solutions explored in this paper, the
requirements that the central gravity dominate the vertical
squeezing and that the IMHD regime be applicable are never
satisfied at the location
of the centrifugal shock: in the
fiducial and strong-gravity IMHD cases, magnetic squeezing
exceeds the tidal force in the central gravity-dominated region, whereas
in the fast-rotation IMHD and AD cases, self-gravity dominates
at the location of the shock.

\section{Incorporation of a Centrifugally Driven Wind into the
Disk Solution}
\label{wind}
As we note in \S~\ref{discussion}, the asymptotic
ambipolar-diffusion disk solution satisfies the launching
condition $B_{r,s}/B_z>1/\sqrt{3}$ for a cold, centrifugally
driven wind, and, in fact, implies a radial scaling of the
magnetic field components (at any given instant of time) that is
identical to that of the time-independent, radially self-similar
wind solution of \citeauthor{Blandford1982}
(\citeyear{Blandford1982}, hereafter BP). Once a
super-Alfv\'enic wind of this type is established,
it constitutes the dominant mechanism of
removing angular momentum from the disk (at least in the
vertical direction). Furthermore, if the outflow is quasi steady
on the accretion time scale, then the spatially self-similar
wind solution can be
used to determine the value of
$B_{\phi,s}$ (or, equivalently, of the parameter $\delta$) in the asymptotic
self-similar (in space and time) disk solution. An explicit wind
solution also makes it possible to evaluate the reduction in the
mass accretion rate
onto the central star caused by the diversion of part of the
inflowing mass into a disk outflow. Motivated by these
considerations, we now show how the centrifugally driven wind
solution can be incorporated into the asymptotic AD disk model.
As demonstrated in \S~\ref{ADstandard}, the numerical solution of the disk
equations typically converges rapidly behind the centrifugal
shock to the asymptotic form given by equations
(\ref{asymptoticm})--(\ref{asymptotich}). This indicates that, if
the model parameters do not strongly differ from the fiducial values,
then the results that we obtain for the asymptotic regime should
be applicable to the bulk of the rotationally supported disk
that forms around the central object.

The mass loss in the BP model is constant
for each decade of disk radius and can be related to the
accretion rate $\dot M_c$ at the inner edge of the disk by
\begin{equation}
\label{mdot_wind}
\frac{1}{2\pi r} \frac{\partial {\dot M_w}}{\partial r}=
\frac{\epsilon \dot M_c}{2\lambda -3} \frac{1}{2\pi r^2}\ ,
\end{equation}
where $\epsilon\leqslant 1$ is the fraction of the disk binding energy at
its inner edge $\rin$ that is carried off by the wind, and
where $\lambda$ is the total (kinetic plus magnetic) specific
angular momentum in the wind, normalized by the Keplerian disk
value $rV_K=(GM_cr)^{1/2}$. The field-line constant $\lambda$ is one
of the three parameters that define a solution in the cold,
self-similar BP wind model, the other two being $\xi_0^\prime
\equiv B_{r,s}/B_z$ and $\kappa \equiv 4\pi \rho V_z V_K/B_z^2$
(the normalized mass-to-magnetic flux ratio evaluated at the
disk surface). As we noted above, for any given value of $t$ in
the asymptotic AD disk solution, $B_{r,s}/B_z$ is a spatial constant
($=4/3$; see eq.\ [\ref{ADpsistandard}]) that exceeds the
launching threshold $1/\sqrt{3}$. It is thus natural to set
$\xi_0^\prime=4/3$. One can then evaluate the parameter
$\kappa$ by equating $2\rho V_z$ with the expression given by
equation (\ref{mdot_wind}) (where the factor of 2 accounts for
the two disk surfaces). The result (in terms of the
nondimensional flow quantities introduced in
\S~\ref{selfsimilareqns}) is
\begin{equation}\label{kappa_1}
\kappa = \frac{\epsilon \dot m_c
m_c^{1/2}}{(2\lambda-3)b_z^2x^{5/2}}\ .
\end{equation}
Substituting for $\dot m_c$ and $b_z$ from the
asymptotic-solution equations (\ref{asymptoticm}) and
(\ref{ADbz}), respectively, this becomes
\begin{equation}\label{kappa_2}
\kappa = \frac{2\delta\epsilon}{2\lambda-3}\ .
\end{equation}
We now substitute for $\delta$ in the asymptotic
disk solution the corresponding expression in the BP wind model,
\begin{equation}\label{delta}
\delta=-\frac{b_{\phi,s}}{b_z}= \kappa(\lambda-1)\, ,
\end{equation}
which yields
\begin{equation}\label{lambda}
\lambda = \frac{3-2\epsilon}{2(1-\epsilon)}\ .
\end{equation}
Although we started by evaluating $\kappa$, we ended up with an
expression that gives $\lambda$ in terms of $\epsilon$. The
latter parameter was treated by
BP as being distinct from $\lambda$, but we
are now able to relate them by combining the disk and wind
solutions. By fixing the values of $\xi_0^\prime$ and
$\lambda$, one can derive the value of $\kappa$ from the
solution of the self-similar wind equations (see Fig.\ 2 in
BP), and thereby obtain $\delta$  (using
eq.\ [\ref{delta}]). In practice, this ``recipe'' for
incorporating the wind model does not alter our basic disk model
since it effectively amounts to replacing one somewhat arbitrary
parameter ($\delta$) by another ($\epsilon$). However, by
combining the disk and wind solutions, it is possible to constrain the
parameter values. In particular, by requiring that the flow
be magnetically dominated in the vicinity of the disk (which
implies $\kappa<1$) and that it become super-Alfv\'enic at a
finite height, one can infer a lower bound on $\lambda$. Based on
Figure 2 in BP, this lower limit is $\sim 3$ for
$\xi_0^\prime=4/3$ (which, by eq.\ [\ref{lambda}], implies that
$\epsilon$ must be $\gtrsim 0.75$), and as $\lambda$ increases from
$\sim 3$ to $\sim 30$ (corresponding to $\epsilon$ increasing from
$\sim 0.75$ to $\sim 0.98$), $\delta$ decreases from $\sim
2$ to $\sim 0.3$.

In the absence of a disk outflow, the central mass eigenvalue $m_c$
is given approximately by the product $-xu\sigma$ evaluated at
$x_c$ (see eqs.\ [\ref{mdot}] and [\ref{asymptoticm}]). When the
mass loss between the outer edge of the disk ($\rout\approx r_c$)
and its inner edge ($\rin\approx R_*$, the protostellar radius)
is taken into account in the mass conservation equation, the value of $m_c$
is reduced to
\begin{equation}\label{m_cw}
m_c \approx \left [1+\frac{\ln{\Lambda}}{2(\lambda-1)}\right
]^{-1}(-xu\sigma)_{x_c}\, ,
\end{equation}
where $\Lambda \equiv \rout/\rin$ (see
eqs.\ [\ref{mass4d}], [\ref{mdot_wind}], and [\ref{lambda}]). To the
extent that deuterium burning causes the radius of the accreting
protostar to increase roughly linearly with its
mass \citep[e.g.,][]{Stahler1988}, $\Lambda$ can be approximated
as a constant. Estimating $r_c$ from our fiducial AD solution
(see eq.\ [\ref{xcADapproximation}]) and $R_*$ from
the mass--radius relation given by \citet{Stahler1988}, we
obtain $\ln{\Lambda} \approx 8$. Equation (\ref{m_cw}) then
implies that mass outflow from the disk
surfaces could reduce the accretion rate onto the central object
by a factor as large as $\sim 3$ (using the minimum
value of $\lambda$, which corresponds to $\kappa=1$). More
strongly magnetized outflows have lower $m_c$ reduction
factors: for example, when $\kappa$ decreases to $\lesssim 0.1$
(corresponding to $\lambda \gtrsim 10$ and $\delta\lesssim 1$), this factor
declines to $\lesssim 1.5$. Although the cumulative effect of
the mass loss from the wind surfaces (embodied in the factor
$\ln{\Lambda}$) can be significant, its impact at any
given radius is comparatively small because of the large range
of radii involved ($\rout/\rin\approx 3\times 10^3$
in the fiducial case). The incorporation of a disk wind modifies the integrated
self-similar disk equations only through logarithmic terms in $x$, so to lowest
order the power-law scalings of the asymptotic disk solution remain unchanged.

Explicit models of AD-dominated, wind-driving accretion disks
were previously constructed by \citet{WK} and
\citet{Li96}. Two general types of solutions were obtained,
depending on the value of the neutral--ion coupling parameter
$\etaWKL \equiv 1/\tni\Omega_K$. Strongly coupled disks have
$\etaWKL \gtrsim 1$ at all heights and are characterized by the
thermal pressure not being much larger
than the magnetic pressure at $z=0$, by
$B_r$ starting to increase already near the midplane and
generally exceeding $|B_\phi|$ at the disk surface (which leads to
a strong magnetic squeezing of the disk), and by the inflow speed
typically being $\gtrsim C$ \citep{WK}. In contrast, in weakly
coupled disks (characterized by $\etaWKL<1$ over the bulk
of the vertical column) the thermal pressure can be much larger than the
magnetic pressure at $z=0$, $B_r$ only starts to grow well above
the midplane (at a height where the coupling parameter finally
rises above 1) and usually does not exceed $|B_\phi|$ at the
surface (which results in the magnetic squeezing remaining
relatively unimportant), and the mass-averaged inflow speed is
typically well below $C$ \citep{Li96}. As we point out in \S~\ref{discussion},
$\etaWKL$ is equal to $2/3\delta$ in
the asymptotic AD disk solution. This result is compatible with
the just-described behavior of $B_{r,s}$ and $|B_{\phi,s}|$:
since $B_{r,s}/B_z =4/3$ in the asymptotic solution and
$\delta=|B_{\phi,s}|/B_z$, the expression for
$\etaWKL$ implies $B_{r,s}/|B_{\phi,s}| = 2\etaWKL$
[consistent with eq.\ (4.5) in \citet{WK} and eq.\ (39)
in \citet{Li96}]. The values of $\delta$ derived from the
BP wind solution indicate that our AD disks are likely to be
weakly coupled. This conclusion is consistent with the
low magnetic-to-thermal pressure ratio implied by the asymptotic
disk solution (it is $\propto r^{1/2}$ as $r\rightarrow 0$ at
any given instant of time; see eqs.\ [\ref{ADsigma}]--[\ref{asymptotich}]),
which corresponds to the situation in weakly coupled
disks. The association with the weakly coupled class of
solutions is further indicated by the unimportance of magnetic
squeezing in the asymptotic expression for the disk
half-thickness (eq.\ [\ref{asymptotich}]) and by the
subsonic inflow speed inferred in the asymptotic regime ($V_r/C\propto r^{1/2}$
as $r\rightarrow 0$ and $t$ is held fixed; see eq.\ [\ref{ADu}]).
The inner regions of real protostellar disks
(on radial scales $\sim 0.1{\range}10\, \au$) are, in fact, expected to
be so weakly ionized over most of their interiors that the
requisite field-line bending for launching a centrifugal wind
can only occur in comparatively thin surface layers
\citep[e.g.,][]{Wardle97}. As was, however, already emphasized
by \citet{Li96}, the properties of such disks and of their
associated outflows depend sensitively on the details of the spatial
profile of $\etaWKL$. Further progress in developing
these models will thus require a good understanding of the
ionization structure in the inner regions of protostellar disks.

\end{document}